\documentclass[5p]{elsarticle}

\usepackage{lineno}
\modulolinenumbers[5]
\usepackage{color}
\usepackage{amssymb}
\usepackage{amsmath}
\usepackage{graphicx}
\usepackage{hyperref}
\usepackage{subfigure}
\usepackage{graphicx, caption}
\usepackage{tabularx}
\usepackage{bm}
\usepackage{bbm}
\usepackage{algorithm}
\usepackage{algorithmic}
\usepackage{lineno} 
\usepackage{hyperref}
\usepackage{url}
\usepackage{array}
\usepackage{makecell}
\usepackage{multirow}
\usepackage{siunitx}
\usepackage{appendix}
\usepackage{balance}

\modulolinenumbers[5]

\newcommand{\argmin}{\operatornamewithlimits{argmin}}

\usepackage{stackengine}

\graphicspath{{./pdf/},{./png}}
\DeclareGraphicsExtensions{.pdf,.png}

\journal{NeuroImage}

\begin{document}

\begin{frontmatter}
\title{Joint super-resolution and synthesis of 1 mm isotropic MP-RAGE volumes from clinical MRI exams with scans of different orientation, resolution and contrast }

\author[cmic,martinos,csail]{Juan Eugenio Iglesias\corref{corresponding author}}       
\cortext[corresponding author]{Corresponding author}
\ead{e.iglesias@ucl.ac.uk}
\author[cmic]{Benjamin Billot} 
\author[cmic]{Ya\"{e}l Balbastre}
\author[martinos,mghrad]{Azadeh Tabari}
\author[martinos,mghrad]{John Conklin}
\author[cmic]{Daniel C. Alexander} 
\author[csail]{Polina Golland}
\author[martinos,mghneuro]{Brian L. Edlow}
\author[martinos]{Bruce Fischl}
\author{for the Alzheimer's Disease Neuroimaging Initiative\fnref{adni}}

\address[cmic]{Centre for Medical Image Computing, Department of Medical Physics and Biomedical Engineering, University College London, UK}
\address[martinos]{Athinoula A. Martinos Center for Biomedical Imaging, Massachusetts General Hospital and Harvard Medical School, Boston, USA}
\address[csail]{Computer Science and Artificial Intelligence Laboratory, Massachusetts Institute of Technology, Boston, USA}
\address[mghrad]{Department of Radiology, Massachusetts General Hospital, Boston, USA}
\address[mghneuro]{Center for Neurotechnology and Neurorecovery, Massachusetts General Hospital, Boston, USA}
\fntext[adni]{Data used in preparation of this article were obtained from the Alzheimer's Disease Neuroimaging Initiative (ADNI) database (\url{http://adni.loni.usc.edu}). As such, the investigators within the ADNI contributed to the design and implementation of ADNI and/or provided data but did not participate in analysis or writing of this report. A complete listing of ADNI investigators can be found at: \url{adni.loni.usc.edu/wp-content/uploads/how_to_apply/ADNI_Acknowledgement_List.pdf}.}

\begin{abstract}

Most existing algorithms for automatic 3D morphometry of human brain MRI scans are designed for data with near-isotropic voxels at approximately 1 mm resolution, and frequently have contrast constraints as well -- typically requiring T1-weighted images (e.g., MP-RAGE scans). This limitation prevents the analysis of millions of  MRI scans acquired with large inter-slice spacing  in clinical settings every year (``thick-slice scans''). In turn, the inability to quantitatively analyze these scans hinders the adoption of quantitative neuroimaging in healthcare, and also precludes research studies that could attain huge sample sizes and hence greatly improve our understanding of the human brain.  Recent advances in convolutional neural networks (CNNs) are producing outstanding results in super-resolution and contrast synthesis of MRI. However, these approaches are very sensitive to the contrast, resolution and orientation of the input images, and thus do not generalize to diverse  clinical acquisition protocols -- even within sites. In this article, we present \emph{SynthSR}, a method to train a CNN that receives one or more thick-slice scans with different contrast, resolution and orientation, and produces an isotropic scan of canonical contrast (typically a 1 mm MP-RAGE). The presented method does not require any preprocessing, e.g., skull stripping or bias field correction. Crucially,  \emph{SynthSR} trains on synthetic input images generated from 3D segmentations, and can thus be used to train CNNs for \emph{any} combination of contrasts, resolutions and orientations without high-resolution training data. We test the images generated with \emph{SynthSR} in an array of common downstream analyses, and show that they can be reliably used for subcortical segmentation and volumetry, image registration (e.g., for tensor-based morphometry), and, if some image quality requirements are met, even cortical thickness morphometry. The source code is publicly available at \url{https://github.com/BBillot/SynthSR}.

\end{abstract}


\end{frontmatter}

\section{Introduction}

\subsection{Motivation}

Magnetic resonance imaging (MRI) has revolutionized research on the human brain, by enabling \emph{in vivo} noninvasive neuroimaging with exquisite and tunable soft-tissue contrast. Quantitative and reproducible analysis of brain scans requires automated algorithms that analyze brain morphometry in 3D, and thus best operate on data with isotropic voxels. Most existing human neuroimaging software requires near-isotropic acquisitions that are commonplace in research. Examples include most of the tools in the major packages that arguably drive the field (FreeSurfer, \citealt{fischl2012freesurfer}; FSL, \citealt{jenkinson2012fsl}; SPM, \citealt{ashburner2012spm}; or AFNI, \citealt{cox1996afni}), e.g., for segmentation \citep{dale1999cortical,fischl2002whole,ashburner2005unified,patenaude2011bayesian} or registration \citep{cox1999real,jenkinson2002improved,ashburner2007fast,andersson2007non,greve2009accurate} of brain MRI scans. Many other popular tools outside these packages also have this near-isotropic resolution requirement, including registrations packages like ANTS \citep{avants2008symmetric}, Elastix \citep{klein2009elastix} or NiftyReg \citep{modat2010fast}; and modern segmentation methods based on convolutional neural networks (CNNs) and particularly the U-net architecture \citep{ronneberger2015u,cciccek20163d},  such as DeepMedic \citep{kamnitsas2017efficient}, DeepNAT \citep{wachinger2018deepnat,roy2019quicknat}, or VoxResNet \citep{chen2018voxresnet}.

Moreover, many of these tools require specific sequences and types of MR contrast to differentiate gray and white matter, such as the ubiquitous MP-RAGE sequence \citep{mugler1990three} and its variants \citep{van2008brain,marques2010mp2rage}. Focusing on a specific MR contrast enables algorithms to be more accurate by learning prior distributions of intensities from labeled training data, but also limits their ability to analyze images with contrasts different from that of the training dataset. Most segmentation methods, with the notable exception of Bayesian algorithms with unsupervised likelihood \citep{van1999automated,ashburner2005unified}, have this MRI contrast requirement, and deviations from the expected intensity profiles (``domain shift'', even within T1-weighted MRI) lead to decreased performance, even with intensity standardization techniques \citep{han2006reliability}. The loss of accuracy due to domain shift is particularly large for CNNs, which are fragile against changes in MRI contrast  (see, e.g., \citealt{jog2019psacnn,pmlr-v121-billot20a}). While classic registration algorithms are contrast agnostic, modern deep learning registration techniques \citep{de2019deep,balakrishnan2019voxelmorph} also require images with  MR contrast similar to that of the scans used in training.

However, MRI scans acquired in the clinic are typically quite different from those obtained as part of research studies. Rather than isotropic volumes, physicians have traditionally preferred a relatively sparse set of images of parallel planes, which reduces the time required for acquisition and visual inspection. Therefore,
clinical MRI exams\footnote{In this article, we use the term ``exam'' to refer to the set of scans acquired during a single MRI session.}
typically comprise several scans acquired with different orientations and (often 2D) pulse sequences, each of which consists of a relatively small set of slices (20-30) with large spacing in between (5-7 mm) and often high in-plane resolution (e.g., 0.5 mm). While morphometry of isotropic scans is also starting to be used in the clinic, quantitative imaging in clinical practice is still in its infancy, and the vast majority of existing clinical MRI scans -- including decades of legacy data -- are highly anisotropic, and thus cannot be reliably analyzed with existing tools.

The inability to analyze clinical data in 3D has deleterious consequences in the clinic and in research. In clinical practice, it precludes: quantitative evaluation of the status of a patient compared to the general population; precise measurement of longitudinal change; and reduction of variability in subjective evaluation due to the positioning of the slices. In research, this inability hinders the analysis of millions of brain scans that are currently stored in picture archiving and communication systems (PACS) around the world. For example, approximately 10 million MRI exams were performed in the US alone in 2019 \citep{oren2019curbing}. These figures are far larger than the sample sizes in neuroimaging research studies, which range from dozens of cases to tens of thousands in the largest meta-analyses, such as those by the ENIGMA consortium \citep{thompson2014enigma}. Computing measurements from clinical scans would thus enable research studies with statistical power levels that are currently unattainable, with large potential for improving our understanding of brain diseases.

\subsection{Related work}

There have been many attempts to bridge the gap between clinical and research scans in medical imaging, mostly based on super-resolution (SR) and synthesis techniques, many of which originated from the computer vision literature. SR seeks to obtain an enhanced, high-resolution (HR) image from an input consisting of one or multiple lower-resolution (LR) frames. Early SR was model-based and relied on multiple LR images of the same scene acquired with slight differences in camera positioning; sub-pixel shifts can then be exploited to estimate the HR image, often in combination with a regularizer, i.e., a prior on the HR image \citep{park2003super}. However, this model-based SR with grid-shifted acquisitions is not feasible in the Fourier-encoded (and hence band-limited) dimensions in MRI \citep{scheffler2002superresolution}.  

Successful SR of MRI has been achieved with machine learning (ML) techniques that do not require handcrafting priors of HR images. Instead, they use a dataset of matching LR-HR images to learn a mapping that enables recovery of HR from LR; training data are often obtained by blurring and subsampling HR images to obtain their LR counterparts. Classical ML methods have long been used to learn this mapping, including non-local patch techniques \citep{manjon2010non}, sparse representations \cite{rueda2013single}, low-rank methods \cite{shi2015lrtv}, canonical correlation analysis \citep{bahrami2016reconstruction},  random forests \citep{alexander2017image}, or sparse coding \citep{huang2017simultaneous}.  

These classical techniques have been superseded by deep CNNs, which have achieved very impressive results. Earlier methods relying on older and simpler architectures from the computer vision literature (e.g., \citealt{pham2017brain}, based on the SRCNN architecture, \citealt{dong2015image}) already surpassed classical techniques by a large margin. Further improvements have been provided by the adoption of more recent developments in CNNs, such as densely connected networks \citep{chen2018brain}, adversarial networks \citep{chen2018efficient}, residual connections \citep{chaudhari2018super}, uncertaintly modeling \citep{tanno2020uncertainty}, or progressive architectures \citep{lyu2020multi}. Importantly, it has been shown that the SR images generated with such deep learning techniques can improve downstream analyses, such as cortical thickness \citep{tian2020improving} or tractography \citep{tanno2020uncertainty}.

Meanwhile, MRI contrast synthesis techniques for brain imaging have followed a path parallel to SR. Early methods used classical ML techniques such as dictionary learning \citep{roy2011compressed}, patch matching \citep{iglesias2013synthesizing}, or random forests \citep{huynh2015estimating}. Since MR contrast synthesis is generally an easier problem than SR, these early methods already achieved competitive results; for example, we have shown that patch-matched synthetic images provide almost identical performance as the ground truth scans in downstream tasks such as registration and segmentation \citep{iglesias2013synthesizing}.  Nevertheless, these methods have been superseded by modern ML techniques based on CNNs, often equipped with adversarial losses \citep{goodfellow2014generative} to preserve finer, higher-frequency detail, as well as cycle consistency \citep{zhu2017unpaired} in order to enable synthesis with unpaired data (e.g., \citealt{chartsias2017multimodal,xiang2018deep,nie2018medical,shin2018medical,dar2019image}). 

While the performance of CNNs in SR and synthesis of MRI is impressive, their adoption in clinical MRI analysis is hindered by the fact that they typically require paired data or, at least, HR images of the target contrast in training. This is an important limitation, as such required training data are most often not available -- particularly since the resolution, contrast and orientations acquired in brain MRI exams vary substantially across sites.
To tackle this problem, classical methods based on probabilistic models have been proposed. For example, \cite{dalca2018medical} used collections of thick-slice scans to build a generative model that they subsequently inverted to fill in the missing information between slices. \cite{brudfors2018mri} also cast SR as an inverse problem, using multi-channel total variation as a prior; this approach has the advantage of not needing access to a collection of scans for training, so it can be immediately used for any new set of input contrasts.  \cite{jog2016self} use Fourier Burst Accumulation \citep{delbracio2015removing}  to super-resolve across slices using the high-resolution information existing within slices (i.e., in plane); as \cite{brudfors2018mri}, this technique can also applied to single images. Unfortunately, the performance of these classical approaches is  lower than that of their CNN counterparts.

The closest works related to the technique proposed in this article are those by \cite{huang2017simultaneous} and \cite{zhao2020smore}. The former presents  ``WEENIE'', a weakly-supervised joint convolutional sparse coding method for joint SR and synthesis of brain MRI. WEENIE combines a small set of image pairs (LR of source domain, HR of target domain) with a larger set of unpaired scans, and uses convolutional sparse coding to learn a representation (a joint dictionary) where the similarity of the feature distributions of the paired and unpaired data is maximized. The main limitation of WEENIE is its need for paired data, even if in a small amount. 
\cite{zhao2020smore} is a deep learning version of  \cite{jog2016self}, which relies of training a CNN with high-resolution slices (blurred along one of the two dimensions), and using it  to super-resolve the imaging volume across slices. While this technique does not require HR training data and can be applied to a single scan, it has two disadvantages compared with the method presented here. First, it is unable to combine the information from multiple scans from the same MRI exam, with different resolution and contrast. And second, integration of MR contrast synthesis into the method is not straightforward. 

\subsection{Contribution}
\label{sec:contrib}

As explained above, the applicability of deep learning SR and synthesis techniques to clinical MRI is often impractical due to substantial differences in MR acquisition protocols across sites. Even within a single site, it is common for brain MRI exams to comprise different sets of sequences -- particularly when considering longitudinal data, since acquisition protocols are frequently updated and improved, and the same patients may be scanned on different platforms (possibly with different field strengths). 

In this article we present \emph{SynthSR},  a solution to this problem that uses synthetically generated images to train a CNN -- an approach that we recently applied with success to contrast-agnostic and partial volume (PV) segmentation of brain MRI  \citep{pmlr-v121-billot20a,billot2020pv}. The synthetic data mimic multi-modal MRI scans with channels of different resolutions and contrasts, and include artifacts such as bias fields, registration errors, and resampling artifacts. Having full control over the generative process allows us to train CNNs for super-resolution, synthesis, or both, for any desired combination of MR contrasts, resolution, and orientation --  without ever observing a real HR scan of the target contrast, thus enabling wide applicability. 

To the best of our knowledge, \emph{SynthSR} is the first deep learning technique that enables ``reconstruction'' of an isotropic scan of a reference MRI contrast from a set of thick-slice scans with different resolutions and pulse sequences. We extensively validate the applicability of our approach by analyzing the performance of common neuroimaging tools on the reconstructed isotropic scans, including: segmentation for volumetry, registration for tensor-based morphometry, and cortical thickness. This approach contrasts with most of the existing SR literature, where validation relies on image similarity metrics (e.g., peak signal-to-noise ratio)  that may not be a good predictor of performance in the downstream analyses that one is ultimately interested in.

The rest of this paper is organized as follows: Section~\ref{sec:methods} describes our proposed framework to generate synthetic images, and how it can be used to train CNNs for SR, synthesis, or both simultaneously. Section~\ref{sec:ExpAndRes} presents three different experiments that evaluate our proposed method with synthetic and real data, and compare its performance with Bayesian approaches. 
Finally, Section~\ref{sec:discussion} discusses the results and concludes the article with a consideration of future directions and applications of this technique.

\begin{figure*}[!th]
\centering
\includegraphics[width=1.0\textwidth, keepaspectratio]{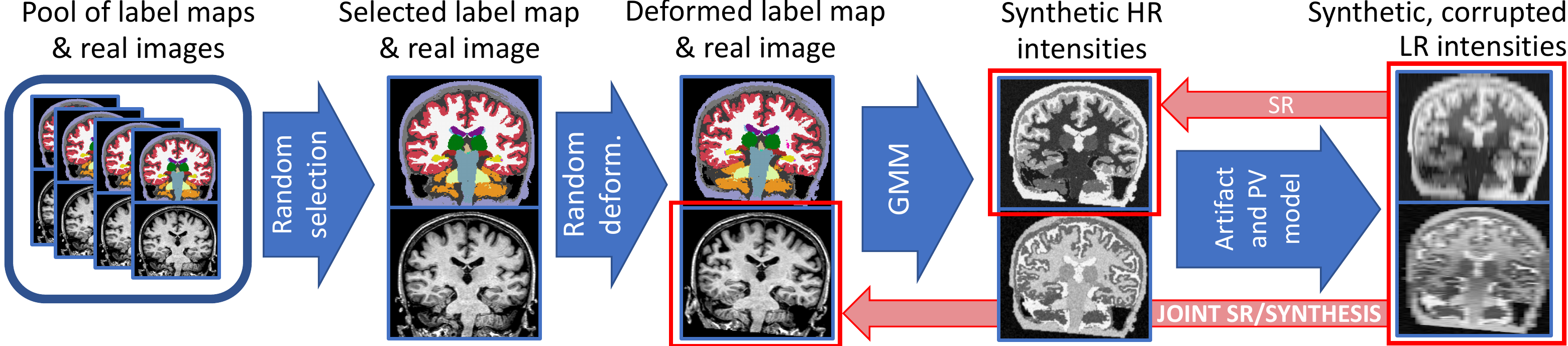}
\captionsetup{width=1.0\textwidth}
\caption{Overview of the synthetic data generator  used by \emph{SynthSR}.  The blue arrows follow the generative model, which is used to sample random scans at every minibatch using a GPU implementation. The red arrows connect the inputs and regression targets used in training for SR or joint SR / synthesis. We emphasize that the real images are only required for joint SR / synthesis, and not SR alone.}
\label{fig:framework}
\end{figure*}

\section{Methods}
\label{sec:methods}

\subsection{Synthetic data generator}
\label{sec:generator}
The cornerstone of \emph{SynthSR} is a synthetic data generator that enables training CNNs for SR and synthesis using brain MRI scans of any resolution and contrast \citep{pmlr-v121-billot20a,billot2020pv}. At every minibatch, this generator is used to randomly sample a series of synthetic images that are used to update the CNN weights via a regression loss. Crucially, this generator is implemented in the GPU, so it does not significantly slow down training. The flowchart of the generator is illustrated in Figure~\ref{fig:framework}; the different steps are described below.

\subsubsection{Sample selection}
For training, we assume the availability of a pool of HR brain scans with the same MR contrast $\{I_n\}_{n=1,\ldots,N}$, together with corresponding segmentations (``label maps'') of $K$ classes  $\{L_n\}_{n=1,\ldots,N}$ corresponding to brain structures and extracerebral regions; these segmentations can be manual, automated, or a combination thereof. Importantly, the MR contrast of these volumes \emph{defines} the reference contrast we will synthesize, so they would typically be 1 mm isotropic MP-RAGE scans; if one wishes to perform SR alone (i.e., without synthesis), these images are not required. At every minibatch, the generative process starts by randomly selecting an image-segmentation pair $(I,L)$ from the pool using a uniform distribution:
\begin{align}
n & \sim \mathcal{U}(1,N), \nonumber \\
I & \leftarrow I_n,  \nonumber \\
L & \leftarrow L_n.  \nonumber 
\end{align}

\subsubsection{Spatial augmentation}
The selected image and segmentation are augmented with a spatial transform $T$, which is the composition  of a linear and nonlinear transform: $T = T_{lin} \circ T_{nonlin}$. The linear component is a combination of three rotations $(\theta_x,\theta_y,\theta_z)$, three scalings $(s_x,s_y,s_z)$ and three shearings $(\phi_x,\phi_y,\phi_z)$, all sampled from uniform distributions (the scalings are sampled in logarithmic domain): 
\begin{align}
&\theta_x \sim \mathcal{U}(a_{rot},b_{rot}),  \  \log s_x \sim \mathcal{U}(a_{sc},b_{sc}),  \   \phi_x \sim \mathcal{U}(a_{sh},b_{sh}),  \nonumber \\
&\theta_y \sim \mathcal{U}(a_{rot},b_{rot}),  \  \log s_y \sim \mathcal{U}(a_{sc},b_{sc}),  \    \phi_y \sim \mathcal{U}(a_{sh},b_{sh}), \nonumber \\
&\theta_z \sim \mathcal{U}(a_{rot},b_{rot}),  \  \log s_z \sim \mathcal{U}(a_{sc},b_{sc}) , \    \phi_z \sim \mathcal{U}(a_{sh},b_{sh}), \nonumber \\
&T_{lin} = \text{Affine}(\theta_x,\theta_y,\theta_z,s_x,s_y,s_z,\phi_x,\phi_y,\phi_z), \label{eq:linAug}
\end{align}
where $a_{rot}, b_{rot}, a_{sc}, b_{sc}, a_{sh}, b_{sh}$ are the minimum and maximum values of the uniform distribution, and  $\text{Affine}(\cdot)$ is an affine matrix consisting of the product of nine matrices: three scalings, three shearings, and three rotations about the $x$, $y$ and $z$ axis. We note that we do not include translation into the model, since it is not helpful in a dense prediction setup -- as opposed to, e.g., image classification.

The nonlinear component is a diffeomorphic transform generated as follows. First, we generate a low dimensional volume with three channels (e.g., $10 \times 10 \times 10 \times 3$) by randomly sampling a zero-mean Gaussian distribution at each location independently. Second, we trilinearly upsample these three channels to the size of the image $I$ in order to obtain a smooth volume with three channels, which we interpret as a stationary velocity field (SVF). Finally, we compute the Lie exponential via integration of the SVF with a scale-and-square approach~\citep{arsigny2006log} in order to obtain a diffemorphic field that is smooth and invertible:
\begin{align}
\text{SVF}' &\sim \mathcal{N}_{10 \times 10 \times 10 \times 3}(0,\sigma^2_{T}), \nonumber \\
\text{SVF} &= \text{Upsample}(\text{SVF}'), \nonumber \\
T_{nonlin} &= \exp(\text{SVF}). \nonumber
\end{align}
where the variance $\sigma^2_{T}$ controls the smoothness of the field. 

Finally, the composite deformation $T$ is used to deform $I$ and $L$ into $I^T$ and $L^T$ using trilinear and nearest neighbor interpolation, respectively:
\begin{align}
I^T  = I \circ T  &= I \circ (T_{lin} \circ T_{nonlin}) \nonumber \\
L^T  = L \circ T  &= L \circ (T_{lin} \circ T_{nonlin}) \nonumber
\end{align}

\subsubsection{Synthetic HR intensities}
\label{sec:HRintensities}
Given the deformed segmentation $L^T$, we subsequently generate HR intensities by sampling a Gaussian mixture model (GMM) at each location, conditioned on the labels. This GMM is in general multivariate (with $C$ different channels corresponding to $C$ MR contrasts) and has as many components as the number of classes $K$. The intensities are further augmented with a random Gamma transform. Specifically, the GMM parameters and HR intensities are randomly sampled as follows:
\begin{align}
\mu_{k,c} \sim & \mathcal{N}(m_{k,c}^{\mu},a_{k,c}^{\mu}), \nonumber \\
\sigma_{k,c} \sim & \mathcal{N}_{trunc}(m_{k,c}^{\sigma},a_{k,c}^{\sigma}), \nonumber \\
G'_c(x,y,z) \sim & \mathcal{N}\left(\mu_{L^T(x,y,z),c},\sigma^2_{L^T(x,y,z),c}\right), \\ \nonumber
\gamma_c \sim & \mathcal{U}(a_\gamma,b_\gamma), \nonumber \\
G_c = & \min_{x,y,z} G'_c + (\max_{x,y,z} G'_c - \min_{x,y,z} G'_c) \times \nonumber \\
& \left[\frac{G'_c -\min_{x,y,z} G'_c}{\max_{x,y,z} G'_c - \min_{x,y,z} G'_c}\right]^{\gamma_c}, \nonumber \\
G(x,y,z) & = \{G_c(x,y,z)\}_{c=1,\ldots,C}, \nonumber
\end{align}
where the mean and standard deviation $(\mu_{k,c},\sigma_{k,c})$ of each class $k$ and MR contrast/channel $c$ are independently sampled from Gaussian distributions (the latter truncated to avoid negative values), and the Gaussian intensity at HR $G_c$ is independently sampled at each spatial location $(x,y,z)$ from the distribution class indexed by the corresponding label $L^T(x,y,z)$. Note that we assume the covariances between the different contrasts to be zero. The hyperparameters $\{m_{k,c}^{\mu}\},\{a_{k,c}^{\mu}\},\{m_{k,c}^{\sigma}\},\{a_{k,c}^{\sigma}\}$ control the contrast of the synthetic images; the practical procedure we follow to estimate these parameters is detailed in Section~\ref{sec:hyperparameters} below. Finally, the parameters $a_\gamma, b_\gamma$ of the uniform distribution for $\gamma$ control the maximum strength of the nonlinear gamma transform. We note that this highly flexible process generates a very wide variety of contrasts -- much wider than what one encounters in practice. Our goal is not to faithfully reproduce the image formation model of MRI, but to generate a diverse set of images, as there is increasing evidence that exposing CNNs to a broader range of images than they will typically encounter at test time improves their generalization ability (see for instance \citealt{chaitanya2019semi}).

\begin{figure}[!t]
\centering
\includegraphics[width=0.5\textwidth, keepaspectratio]{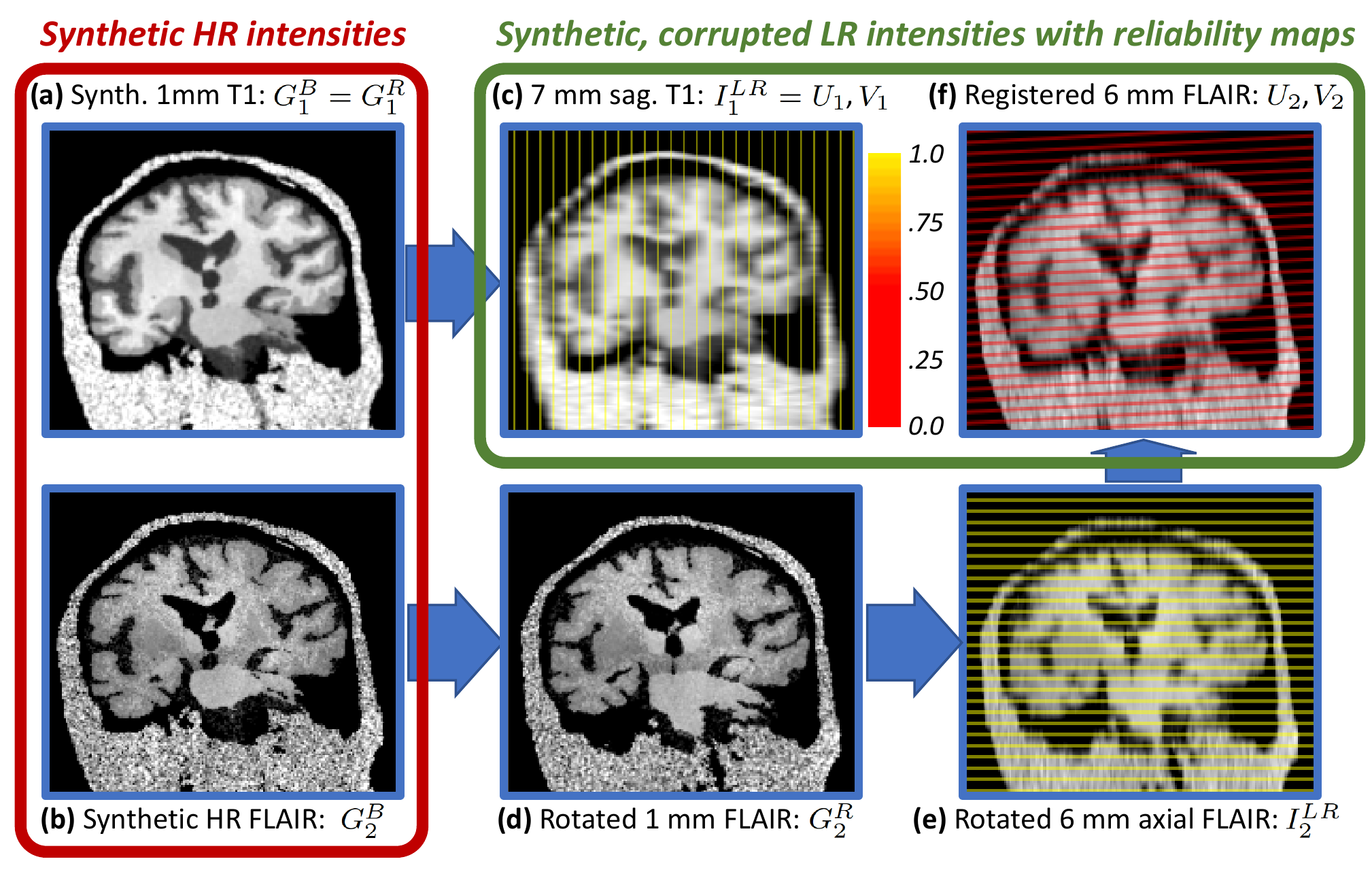}
\captionsetup{width=0.5\textwidth}
\caption{Details of the workflow for the generator of synthetic scans with reliability maps, using an example with a 7 mm sagittal T1 acquisition (used as reference) and a 6 mm axial FLAIR. (a)~Synthetic HR T1 with bias field ($G^B_1 = G_1^R$). (b)~Synthetic HR FLAIR with bias field ($G^B_2$). (c)~Synthetic LR sagittal T1 with reliability map overlaid ($I_1^{LR}=U_1$ and $V_1$). (d)~Synthetic HR FLAIR with small random deformation, simulating subject motion in between scans ($G^R_2$). (e)~Synthetic LR axial FLAIR with reliability map ($I_2^{LR}$). (f)~LR FLAIR and reliability map registered to the reference space defined by the T1 scan ($U_2$ and $V_2$); note that the reliability map is no longer binary or parallel to the axial plane. Registration errors are modeled by adding noise to the inverse of the random rigid transform when deforming back to the reference space.}
\label{fig:frameworkIntensities}
\end{figure}

\subsubsection{Synthetic, corrupted LR intensities}
\label{sec:LRintensities}
The last step of the synthetic data generation is the simulation of variability in coordinate frames and of image artifacts, including bias field, PV, registration errors, and resampling artifacts.

\paragraph{Variability in coordinate frames}
In practice, the different channels of multi-modal MRI scans are  not perfectly aligned due to inter-scan motion, i.e., the fact that subject moves in between scans. Therefore, a first step when processing data from an MRI exam is to select one of the input channels to define a reference coordinate frame, and register all the other channels to it. Inter-scan motion aside, the coordinate frames of the different channels are in general not perfectly orthogonal, for two possible reasons. First, it is possible that the geometric planning of the different channels is not orthogonal by design. For example, the coronal hippocampal subfield T2 acquisition in ADNI is oriented perpendicularly to the major axis of the hippocampus, and is thus rotated with respect to the isotropic 1mm MP-RAGE acquisition. And second, the aforementioned inter-scan motion. In order to model these differences, we apply random rigid transforms to all the MR contrasts except for the reference channel, which we assume, without loss of generality, to be the first one:
\begin{align}
&\theta^R_{c,x} \sim \mathcal{U}(a_{rot},b_{rot}),  \   t_{c,x} \sim \mathcal{U}(a_t,b_t),   \nonumber \\
&\theta^R_{c,y} \sim \mathcal{U}(a_{rot},b_{rot}),  \   t_{c,y} \sim \mathcal{U}(a_t,b_t), \nonumber \\
&\theta^R_{c,z} \sim \mathcal{U}(a_{rot},b_{rot}),  \   t_{c,z} \sim \mathcal{U}(a_t,b_t), \nonumber \\
&R_c  = 
    \begin{cases}
      \text{Id.} = \text{Rigid}(0,0,0,0,0,0),  &  \text{if}\ c=1\\
        \text{Rigid}(\theta^R_{c,x},\theta^R_{c,y},\theta^R_{c,z},t_{c,x},t_{c,y},t_{c,z}),  & \text{if}\ c>1
    \end{cases} \label{eq:rigidTransform} \\
&G^R_c  =  G_c \circ R_c,  \nonumber
\end{align}
where we use the same parameters of the uniform distribution of the rotation angles as in Equation~\ref{eq:linAug}, $a_t, b_t$ are the extremes of the uniform distribution for the translations, $\text{Rigid}(\cdot)$ is a rigid transform matrix consisting of the product of three rotation and three translation matrices,  $R_c$ is the rigid transformation matrix for channel $c$, and $G_c^R$ is the rigidly deformed synthetic HR volume for contrast $c$. An example of this deformation is shown in Figure~\ref{fig:frameworkIntensities}(d). 

\paragraph{Bias field}
In order to generate a smooth multiplicative bias field, we use a strategy very similar to the one we utilized for the nonlinear deformation, and which consists of four steps that are independently repeated for each MR contrast $c$. First, we generate a low dimensional volume  (e.g.,  $4 \times 4 \times 4$)  by randomly sampling a zero-mean Gaussian distribution at each location independently. Second, we linearly upsample this volume to the size of the full image  $G_c$. Third, we take the voxel-wise exponential of the volume to obtain the bias field $B_c$. And fourth, we multiply each channel $c$ of the Gaussian volume $G_c$ by  $B_c$ at every spatial location:
\begin{align}
\log B'_c &\sim \mathcal{N}_{4 \times 4 \times 4 }(0,\sigma^2_{B}), \nonumber \\
\log B_c &= \text{Upsample}(\log B'_c), \nonumber \\
B_c(x,y,z) &= \exp [\log B_c(x,y,z)], \nonumber \\
G^B_c(x,y,z) &= G^R_c(x,y,z) B_c(x,y,z), \nonumber
\end{align}
where the variance $\sigma^2_{B}$ controls the strength of the bias field, $B_c(x,y,z)$  the non-negative bias field at location $(x,y,z)$, and $G^B_c(x,y,z)$ represents the corrupted intensities of (rigidly deformed) channel $c$.

\paragraph{Partial voluming}
The simulation of partial voluming happens independently for every channel, and has two aspects: slice thickness and  slice spacing. The slice thickness can be simulated by blurring in the corresponding direction. The blurring kernel is directly related to the MRI slice excitation profile. While real sequences use complicated profiles (e.g., numerically optimized with the Shinnar-Le Roux algorithm, \citealt{pauly1991parameter}), we instead use a simple Gaussian kernel in our simulations. The standard deviations of the kernels $\bm{\sigma}_{S,c}$ depend on the direction and channel, and are designed to divide the power of the HR signal by 10 at the cut-off frequency \citep{billot2020pv}. We further multiply $\bm{\sigma}_{S,c}$ by a random factor $\alpha$, sampled from a uniform distribution of predefined range, to model small deviations from the nominal voxel size.  

Once the image has been blurred, slice spacing can be easily modeled by subsampling every channel in every direction with the prescribed channel-specific spacing distances. This subsampling produces synthetic, corrupted, misaligned LR intensities for every channel $c$. The specific processing is:
\begin{align}
\alpha &\sim \mathcal{U}(a_{\alpha}, b_{\alpha}), \nonumber \\
\bm{\sigma}_{S,c}  &= 2 \alpha \log(10) / (2\pi)   \bm{r}_c / \bm{r}_{targ} ,  \nonumber \\
I_c^{\sigma} &= G^R_c \ast \mathcal{N}[0,diag(\bm{\sigma}_{S,c})], \nonumber \\
I_c^{LR} &= \text{Resample} \left( I_c^{\sigma} ; \bm{d}_c \right), \nonumber 
\end{align}
where $a_{\alpha}, b_{\alpha}$ are the parameters of the uniform prior distribution  over $\alpha$; $\bm{r}_c$ is the (possibly anisotropic) voxel size  of the test scan in channel $c$, without considering gaps between slices; $\bm{r}_{targ}$ is the (often isotropic) voxel size of the training segmentations (which defines the target resolution for SR); $I_c^\sigma$ is the blurred channel $c$;  $\text{Resample}(\cdot)$ is the resample operator;   $\bm{d}_c$ is the voxel spacing of channel $c$; and $I_c^{LR}$ are the synthetic, corrupted, misaligned LR intensities.  We note that $\bm{r}_c$, $\bm{r}_{targ}$ and  $\bm{\sigma}_{S,c}$ are $3 \times 1$ vectors, with components for the $x$, $y$ and $z$ directions. Examples of PV modeling are shown in Figure~\ref{fig:frameworkIntensities}(c,e).

\paragraph{Registration errors and resampling artifacts}
The final step of our generator is mimicking the preprocessing that will happen at test time, where the different channels will be rigidly registered to the reference channel $c=1$ and trilinearly upsampled to the (typically isotropic) target resolution $\bm{r}_{targ}$. At that point, all images are defined on the same voxel space, and SR and synthesis become a voxel-wise regression problem. In order to simulate the registration step, one could simply invert the rigid transform modeling the variability in coordinate frames (Equation~\ref{eq:rigidTransform}). However, registration will always be imperfect at test time, so it is crucial to simulate registration errors in our generator. The final images produced of the generator $\{U_c\}$ are given by:
\begin{align}
&\bm{\epsilon}_c  \sim
    \begin{cases}
    \delta(\bm{\epsilon}_c), & c=1 \\
    \mathcal{N}[0, \text{diag}(\sigma^2_{\epsilon,\theta},\sigma^2_{\epsilon,\theta},\sigma^2_{\epsilon,\theta},\sigma^2_{\epsilon,t},\sigma^2_{\epsilon,t},\sigma^2_{\epsilon,t})] ,  &  c>1
     \end{cases}  \nonumber \\
&R'_c= R_c^{-1} \times  \text{Rigid}(\bm{\epsilon}_c), \nonumber \\
&U'_c =  I_c^{LR} \circ R'_c, \nonumber \\
&U_c = \text{Resample}(U'_c; \bm{r}_{targ}), \label{eq:resampling}
 \end{align}
where $\delta(\cdot)$ is Kronecker's delta and $\sigma^2_{\epsilon,\theta},\sigma^2_{\epsilon,t}$ are the variances of the rotation and translation components of the registration error, which are assumed to be statistically independent. An example of a registered and resampled image is shown in Figure~\ref{fig:frameworkIntensities}(f), where the rotation has introduced noticeable resampling artifacts.

In addition to $\{U_c\}$, the generator also produces a second set of volumes $\{V_c\}_{c=1,\ldots,C}$  that we call ``reliability maps''. The reliability maps  encode the confidence one has in the observations at each voxel location and improve the performance of the CNN in practice. They are essentially soft, voxel-wise maps indicating whether voxels are measured or interpolated. Voxels on slices of $I_c^{LR}$ have reliability one, whereas voxels between slices have reliability zero -- see for instance Figure~\ref{fig:frameworkIntensities}(c,e). Reliabilities between zero and one are obtained due to linear interpolation when the target resolution is not an exact multiple of the slice spacing, or when applying the transformation $R'_c$ (simulating the registration) to the maps in order to bring them into alignment with $\{U_c\}$, e.g., as in Figure~\ref{fig:frameworkIntensities}(f). We note that these maps are known for every image, and we use them as additional input at testing (Section~\ref{sec:inference}).

\begin{table*}[!th]
\centering
 \begin{tabular}{|c|c|c|c|c|c|c|c|c|c|c|c|c|c|c|c|}
  \hline
 $a_{rot}$ & $b_{rot}$ & $a_{sc}$ & $b_{sc}$ & $a_{sh}$ & $b_{sh}$ & $\sigma_T^2$ & $a_\gamma$ & $b_\gamma$&  $\sigma^2_B$ & $a_b$ & $b_t$ & $a_\alpha$ & $b_\alpha$ & $\sigma^2_{\epsilon,\theta}$ & $\sigma^2_{\epsilon,t}$ \\
  \hline
  -10 & 10 & 0.9 & 1.1 & -0.01 & 0.01 &  $3^2$ & 0.7 & 1.3 &  $0.5^2$ & -20 & 20  & 0.8 & 1.2 & $0.3^2$ & $0.3^2$\\
  \hline
  \end{tabular}
\captionsetup{width=1.0\textwidth}
\caption{Model hyperparameters. Angles are in degrees, and spatial measures are in mm. }
\label{tab:hyperparameters}
\end{table*}

\subsection{Learning and inference}

\subsubsection{Regression targets and loss}
\label{sec:regressionTarget}
We train a CNN to predict the desired output $Y$ from the inputs $\{U_c, V_c\}$, i.e., the registered LR scans resampled at $\bm{r}_{targ}$ and their corresponding reliability maps, which are generated on the fly during training. We consider two different modes of operation: SR alone, and joint SR and synthesis (Figure~\ref{fig:framework}). In the former case, we seek to recover the synthetic HR volume of the reference contrast $G_1^B = G_1^R$. Rather than predicting this image volume directly, it is an easier optimization problem to predict the residual instead, i.e., we seek to regress $Y = G_1^B - U_1$ from $\{U_c, V_c\}$. This mode of operation does not require any real images for training. 

In joint SR / synthesis, we instead seek to recover the real image intensities of standard contrast, typically MP-RAGE. If any of the input contrasts $c^*$ is similar to the target standard contrast (e.g., a T1-weighted scan acquired with a TSE sequence), we regress the residual, as in the SR case: $Y = I^T - U_{c^*}$. If not, we simply regress the target intensities directly: $Y = I^T$.

The CNN is trained with the Adam optimizer \citep{kingma2014adam}, seeking to minimize the expectation of the L1 norm of the error:
$$
\hat{\Omega} = \argmin_{\Omega}  \mathrm{E} [  \|  Y - \tilde{Y}(U_1, V_1, \ldots, U_c,V_c; \Omega)  \|_1  ]
$$
where $\Omega$ is the set of CNN weights, and $\tilde{Y}(\cdot,\Omega)$ is the output of the CNN when parameterized by $\Omega$. The choice of the L1 norm as loss was motivated by the fact that it produced visually more realistic results in pilot experiments compared with the L2 norm or structural similarity \citep{wang2004image}.

\subsubsection{Network architecture}

Our CNN builds on an architecture that we have successfully used in our previous work with synthetic MRI scans \citep{pmlr-v121-billot20a,billot2020pv}. It is a 3D U-net \citep{ronneberger2015u,cciccek20163d} with 5 levels. Levels consist of two layers, each of which comprises convolutions with (3$\times$3$\times$3 kernels) and a nonlinear ELU activation \citep{clevert_fast_2016}). The first layer has 24 kernels (i.e., features); the number of features is double after each max-pooling, and halved after each upsampling. The last layer uses a linear activation to produce an estimate of $Y$. The U-net is concatenated with the synthetic data generator into a single model entirely implemented on the GPU, using Keras~\citep{chollet2015keras} with a Tensorflow backend~\citep{abadi_tensorflow_2016}.

 \subsubsection{Hyperparameters}
 \label{sec:hyperparameters}
The generator described in Section~\ref{sec:generator} has a number of hyperparameters, which control the variability of the synthetic scans, in terms of both shape and appearance. Table~\ref{tab:hyperparameters} summarizes the values of the hyperparameters related to shape, bias field, gamma augmentation, variability in coordinate frames and slice thickness, and misregistration. These hyperparameters were set via visual inspection of the output, such that the generator yields a wide distribution of shapes, artifacts and intensity profiles during training -- which increases the robustness of the CNN. Furthermore, we used the same values that provided good performance in previous work \citep{pmlr-v121-billot20a,billot2020pv}.

The hyperparameters that control the GMM parameters $\{m_{k,c}^{\mu}\},\{a_{k,c}^{\mu}\},\{m_{k,c}^{\sigma}\},\{a_{k,c}^{\sigma}\}$ do not have predefined values, since they depend on the MR contrast -- and to less extent, the resolution -- of the dataset that we seek to super-resolve. For every experimental setup, we estimate them with the following procedure. First, we run our Bayesian, sequence-adaptive segmentation algorithm (SAMSEG, \citealt{puonti2016fast}) on a small set of scans from the dataset to segment. Even though the quality of these segmentations is often  low due to PV, we can still use them to compute rough estimates of the mean and variance of the intensities of each class with robust statistics. Specifically, we compute the median as an estimate for $\{\mu_{k,c}\}$, and the median absolute deviation (multiplied by 1.4826, \citealt{leys2013detecting}) as an estimate for $\{\sigma_{k,c}\}$. We then scale the estimated variances by the ratio between the volumes of the HR and LR voxels for every modality, i.e., $(\mathbbm{1}^T \bm{r}_c) / (\mathbbm{1}^T \bm{r}_{targ})$ (where $\mathbbm{1}$ is the all ones vector), such that the blurring operator yields the desired variance in the synthetic LR images. Finally, we fit a Gaussian distribution to each of the means and variances (a truncated Gaussian for the latter, in order to avoid non-negative variances) to obtain $\{m_{k,c}^{\mu}\},\{a_{k,c}^{\mu}\},\{m_{k,c}^{\sigma}\},\{a_{k,c}^{\sigma}\}$. Crucially, we multiply  $\{a_{k,c}^{\mu}\}$ and $\{a_{k,c}^{\sigma}\}$ by a factor of 5 in order to provide the CNN with a significantly wider range of images than we expect it to see at test time, thus making it resilient to variations in acquisition (as already explained in Section~\ref{sec:HRintensities} above), as well as for alleviating segmentation errors made by SAMSEG.

\subsection{Inference}
\label{sec:inference}
At testing, one simply strips the generator from the trained model, and feeds the preprocessed images to super-resolve  $\{U_c\}$ together with the corresponding reliability maps $\{V_c\}$. The process to obtain these preprocessed images is the same as in Section~\ref{sec:LRintensities} above. The first step is to resample all the scans to the target resolution $\bm{r}_{targ}$, while computing the corresponding reliability maps. For the reference channel $c=1$, the resampled scan and its associated reliability map immediately correspond to  $U_1$ and $V_1$, respectively. The other channels $c>1$  need to be rigidly registered; the warped resampled images and reliability maps become $\{U_c\}_{c=2,\ldots,C}$ and $\{V_c\}_{c=2,\ldots,C}$, respectively. In our implementation, we use an inter-modality registration tool based on mutual information and block matching (\citealt{modat2014global}, implemented in the NiftyReg package) to estimate the rigid alignments.

\subsection{Other practical considerations}

\paragraph{Further blurring of synthetic HR images in training}
 In practice, we slightly blur the synthetic HR volumes $\{G^B_c\}$ with a Gaussian kernel with 0.5 mm standard deviation \citep{pmlr-v121-billot20a}; this operation introduces a small degree of spatial correlation in the images, making them look more realistic. This strategy produces slightly more visually appealing results in the purely SR mode, as these synthetic HR images are the target of the regression, but does not affect the output when jointly performing SR and synthesis. 

\paragraph{Normalization of image intensities}
Both during training and at testing, we min-max normalize the input volumes to the interval [0,1]. In training, the normalization depends whether synthesis is being performed or not. In the purely SR mode, the target volume is normalized exactly the same way as the input, in order to keep the residual centered around zero. In the joint SR / synthesis mode, the targets are normalized by scaling the intensities such that the median intensity of the white matter is one. 

\paragraph{Computational burden}
We randomly crop the images during training to $192 \times 192 \times 192$ volumes, which enables training on a 16GB GPU. We set the learning rate to $10^{-4}$, and train the CNNs for 200,000 iterations, which was sufficient for convergence in all our experiments -- there was minimal change in the loss and no perceptible difference in the outputs after approximately 100,000 - 150,000 iterations. Training takes approximately 12 days on a Tesla P100 GPU. Inference, on the other hand, takes approximately three seconds on the same GPU.

\section{Experiments and results}
\label{sec:ExpAndRes}

This section presents three sets of experiments seeking to validate different aspects of \emph{SynthSR}. 
First, we use a controlled setup with synthetically downsampled MP-RAGE scans from ADNI, in order to assess the SR ability of the method on a single volume, as a function of slice spacing. 
In the second experiment, we  test the performance of the method in a joint SR / synthesis task, seeking to turn thick-slice FLAIRs from ADNI into 1 mm MP-RAGEs.
In the third and final experiment, we  apply \emph{SynthSR} to multimodal MRI exams  from Massachusetts General Hospital (MGH), seeking to recover a 1 mm MP-RAGE from a set of different thick-slice sequences.

\subsection{MRI data}

We used three different datasets in this study; one for training, and two for testing. 

\paragraph{Training dataset}
The first dataset, which we used for training purposes in all experiments, consists of 39 T1-weighted MRI scans and corresponding segmentations. The scans were acquired on a 1.5T Siemens scanner with an MP-RAGE sequence at 1 mm resolution, with the following parameters: TR=9.7 ms, TE=4ms, TI=20 ms, flip angle=$10^\circ$. This is the dataset that was used to build the probabilistic atlas for the segmentation routines of FreeSurfer \citep{fischl2002whole}. The segmentations comprise a set of manual delineations for 36 brain MRI structures (the same as in \citealt{fischl2002whole}), augmented with labels for extracerebral classes (skull, soft extracerebral tissue, fluid inside the eyes) automatically estimated with a GMM approach. Modeling of extracerebral tissues enables the application of our method to unpreprocessed images, i.e., without skull stripping. 

\paragraph{ADNI}
The second dataset is a subset of 100 subjects from 
Alzheimer's Disease Neuroimaging Initiative (ADNI\footnote{The ADNI was launched in 2003 by the National Institute on Aging, the National Institute of Biomedical Imaging and Bioengineering, the Food and Drug Administration, private pharmaceutical companies and non-profit organizations, as a \$60 million, 5-year public-private partnership. The main goal of ADNI is to test whether MRI, positron emission tomography (PET), other biological markers, and clinical and neuropsychological assessment can be combined to analyze the progression of mild cognitive impairment (MCI) and early AD. Markers of early AD progression can aid in the development of new treatments and monitor their effectiveness, as well as decrease the time and cost of clinical trials. The Principal Investigator of this initiative is Michael W. Weiner, MD, VA Medical Center and University of California - San Francisco. ADNI has been followed by ADNI-GO and ADNI-2. These three protocols have recruited over 1,500 adults (ages 55-90) from over 50 sites across the U.S. and Canada to participate in the study, consisting of cognitively normal older individuals, people with early or late MCI, and people with early AD. Subjects originally recruited for ADNI-1 and ADNI-GO had the option to be followed in ADNI-2.}), 
50 of them diagnosed with Alzheimer's disease (AD, aged 73.7$\pm$7.3 years), and 50 elderly controls (aged 72.2$\pm$7.9); 47 subject were males, and 53 females. We believe that $n=100$ is a sample size that is representative of many neuroimaging studies, and comparing AD with controls yields well-known volumetric effects that we seek to reproduce with thick-slice scans. We used two different sets of images: T1 MP-RAGE scans with approximately 1 mm isotropic resolution, and axial FLAIR scans with 5 mm slice thickness and spacing. Even though no manual delineations are available for this dataset, we use automated segmentations of brain structures computed with FreeSurfer 7 (and their associated volumes) as a reference standard in our experiments.

\paragraph{MGH}
The third and final dataset consists of 40 subjects scanned at MGH  (18 males, 22 females, aged 55.1$\pm$19.0 years). Cases with large abnormalities, such as tumors or resection cavities, were excluded. The scans were downloaded from the MGH PACS and anonymized in accordance with an IRB-approved protocol, for which informed consent was waived. We selected a subset of four sequences that are acquired for most patients scanned at MGH over the last decade (including these 40): sagittal T1-weighted TSE (5 mm spacing, 4 mm thickness), axial T2-weighted TSE (6 mm spacing, 5 mm thickness), axial FLAIR turbo inversion recovery (6 mm spacing, 5 mm thickness), and 1.6 mm T1 spoiled gradient recalled (SPGR). We emphasize that, despite its apparently high spatial resolution,  the SPGR sequence has a short acquisition time (14 seconds), short TR/TE (3.15/1.37 ms),  partial Fourier acquisition (6/8), and aggressive parallel imaging (GRAPPA with a factor of 3). These parameters lead to relatively blurry images with low contrast-to-noise ratio,  which do not yield accurate measurements, e.g., when analyzed with FreeSurfer -- as we show in the results below.  No manual delineations are available for this dataset, and reliable automated segmentations are not available due to the lack of higher resolution companion scans. 


\subsection{Competing methods}


As mentioned in Section~\ref{sec:contrib}, there are -- to the best of our knowledge --  no joint SR / synthesis methods available for single scans that adapt to MRI contrast, and which can thus be applied without the availability of a training dataset. In this scenario, we use SAMSEG as a competing method. Even though SAMSEG does not provide synthesis or SR, it provides segmentations for scans of any resolution and contrast, which we can use for indirect validation (e.g., ability to detect effects of disease). In the experiments with the MGH dataset, for which multiple scans of the same exam are available (including one with T1 contrast), we compare our method against \cite{brudfors2018mri} -- which is the only available method that we know of that can readily super-resolve a set of volumes of arbitrary contrast into a HR scan.


\subsection{Experiments}

\subsubsection{Super-resolution of synthetically downsampled scans}

Our first experiment seeks to assess the SR capabilities of \emph{SynthSR} as a function of the resolution of the input. To do so, we artificially downsampled the MP-RAGE scans from the ADNI dataset to simulate 3, 5 and 7 mm coronal slice spacing, with 3 mm slice thickness in all cases. We then used our method to predict the residual between the HR images and the (upsampled) LR volumes, without any synthesis -- such that training relies solely on synthetic data, as explained in Section~\ref{sec:regressionTarget}. 

\begin{figure}[!t]
\centering
\includegraphics[width=0.4\textwidth, keepaspectratio]{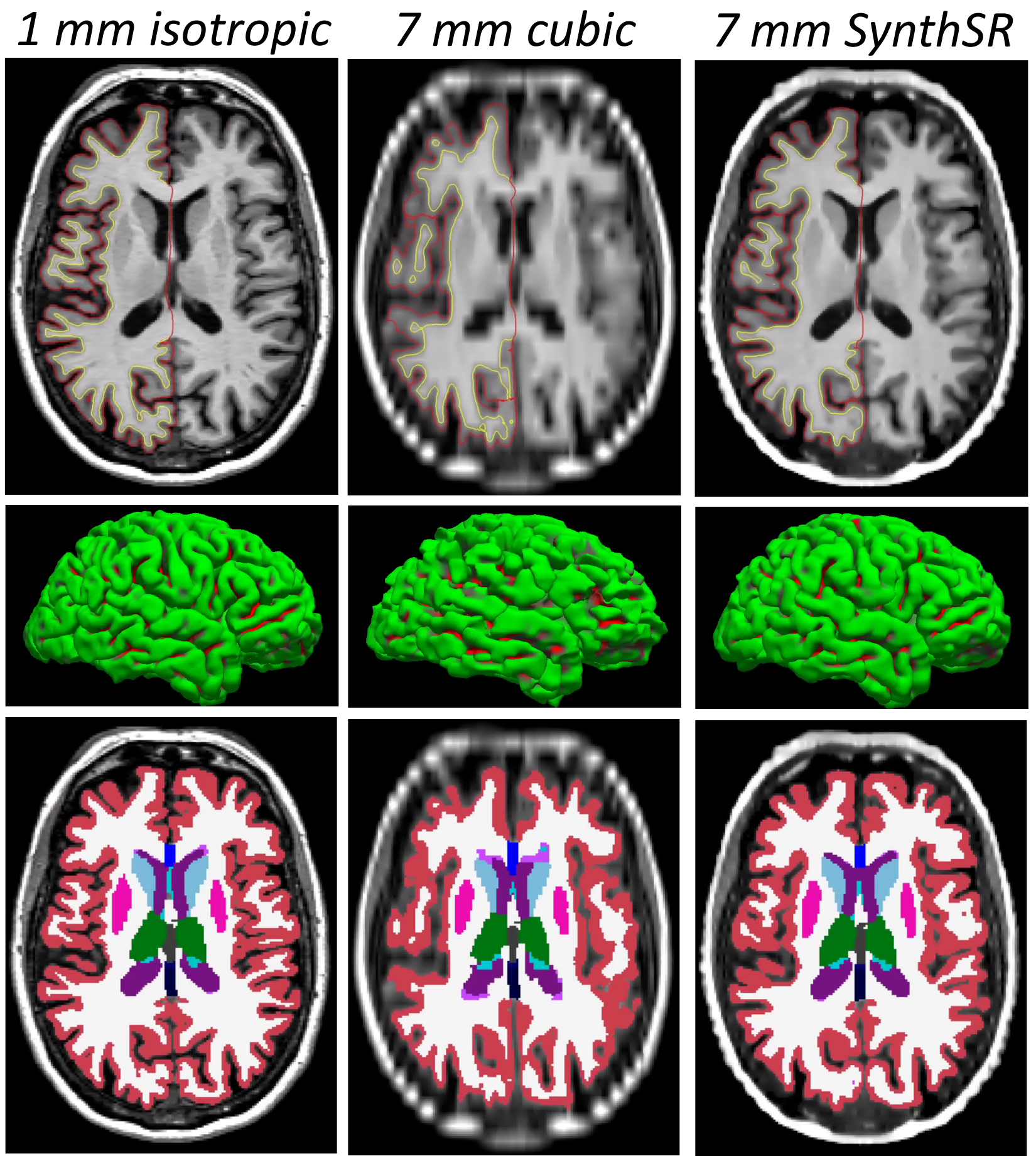}
\captionsetup{width=0.45\textwidth}
\caption{Axial slice of a sample 1 mm T1 scan from the ADNI dataset (left); 7 mm coronal version (middle); and super-resolved back to 1 mm with \emph{SynthSR} (right). Top row: image intensities with pial and white matter surfaces for the right hemisphere (FreeSurfer 7). Middle row: 3D rendered pial surface. Bottom: volumetric FreeSurfer segmentation, represented with the standard FreeSurfer color map.}
\label{fig:upsampling}
\end{figure}

Figure~\ref{fig:upsampling} shows qualitative results for a sample 7 mm scan (1 mm original, downsampled, and super-resolved with \emph{SynthSR}), along with segmentations produced by FreeSurfer~7. Even though the CNN has never been exposed to a real scan during training, \emph{SynthSR} is able to accurately recover high-resolution features; only minimal blurring remains in the SR volume, compared with the original scan. When the 7 mm scan in Figure~\ref{fig:upsampling} is processed directly with FreeSurfer 7 using cubic interpolation, most folding patterns are lost. However, most of these patterns are recovered when the SR volume is processed instead. Subcortically, the segmentation of the LR scan suffers from heavy shape distortion and PV effects (e.g., peri-ventricular voxels segmented as white matter lesions, in lilac), while the SR scan yields a segmentation almost identical to the original.

\begin{table}[!t]
\centering
\begin{small}
 \begin{tabular}{|l|c|c|c|}
  \hline
 Resolution & \makecell{ Average \\ volume error } &  \makecell{ Correlation \\ with 1 mm} &  \makecell{ Effect \\ size}   \\
  \hline
  1 mm & N/A & N/A & 1.38 \\
  \hline
  3 mm (cubic) & 4.5\%   & 0.98   & 1.35    \\
  3 mm (\emph{SynthSR}) & 3.3\%   & 0.99    & 1.36    \\
  5 mm (cubic) & 7.6\%   & 0.95   & 1.22   \\
  5 mm (\emph{SynthSR})  &  2.9\%  & 0.99    & 1.33    \\
  7 mm (cubic) & 10.1\%   & 0.91    &  0.98   \\
  7 mm (\emph{SynthSR}) & 3.0\%  & 0.97   &  1.30     \\
  \hline
  \end{tabular}
  \end{small}
\captionsetup{width=0.45\textwidth}
\caption{Relative error in hippocampal volumes, correlation with ground truth (i.e., FreeSurfer volumes from 1 mm scans), and effect size of AD vs controls (corrected for sex, intracranial volume and age), as a function of slice spacing, with cubic interpolation and \emph{SynthSR}.}
\label{tab:hippoVol}
\end{table}

\begin{figure}[!t]
\centering
\includegraphics[width=0.4\textwidth, keepaspectratio]{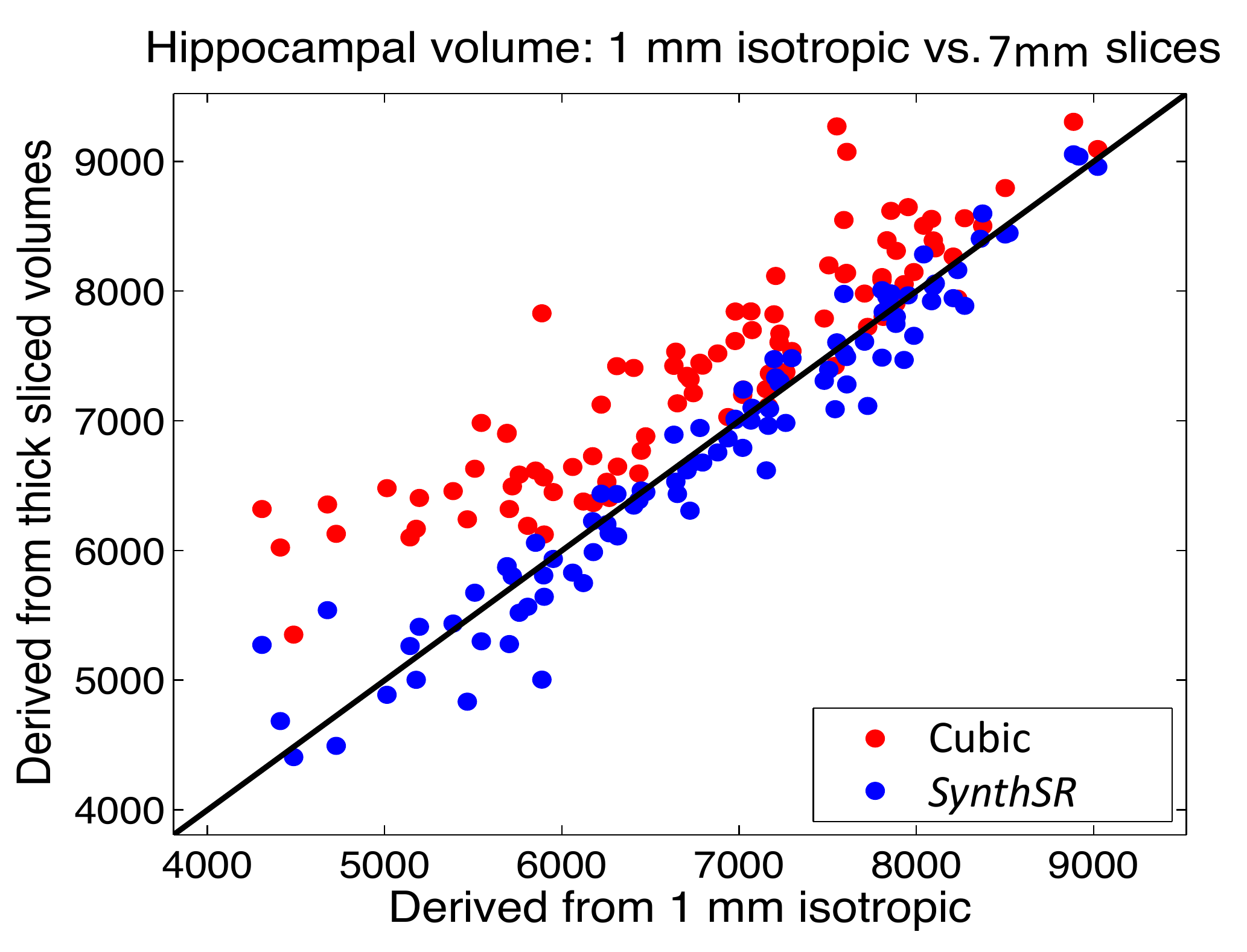}
\captionsetup{width=0.45\textwidth}
\caption{Scatter plot comparing the hippocampal volumes from the 7 mm scans vs the ground truth, using cubic interpolation and \emph{SynthSR}.}
\label{fig:hippoVol}
\end{figure}

In order to assess the performance of \emph{SynthSR} in a quantitative fashion, we evaluate its ability to detect differences between AD and controls through three standard analyses: hippocampal volumetry, cortical thickness, and tensor-based morphometry (TBM). 

\paragraph{Hippocampal volumetry}
Hippocampal volume is a well-known imaging biomarker for AD \citep{gosche2002hippocampal,chupin2009fully,schuff2009mri,shi2009hippocampal}. 
Table~\ref{tab:hippoVol} compares the bilateral hippocampal volume of the AD and control subjects in our ADNI dataset using estimates of the volumes computed with FreeSurfer 7 on the 3, 5 and 7 mm scans, with and without SR, using the volumes from the 1 mm isotropic scans as ground truth. Without SR (i.e., just cubic interpolation), errors grow quickly with slice spacing, while SR with  \emph{SynthSR} keeps the volume errors under 3.5\%, correlations over 0.97, and effect sizes (AD vs. controls, correcting for intracranial volume, sex and age) over 1.30 even for 7 mm spacing  -- compared with 1.38 at 1 mm. The improvement with respect to the non-SR is further illustrated in the scatter plot in Figure~\ref{fig:hippoVol}, which compares the hippocampal volumes from the 1 mm scans (i.e., the reference), with those from the 7 mm scans. Without SR, hippocampal volumes are generally overestimated, particularly for cases with lower volumes, i.e., severe hippocampal atrophy.  \emph{SynthSR}, on the other hand, consistently agrees with the reference across the whole range. 

\begin{figure}[!t]
\centering
\includegraphics[width=0.5\textwidth, keepaspectratio]{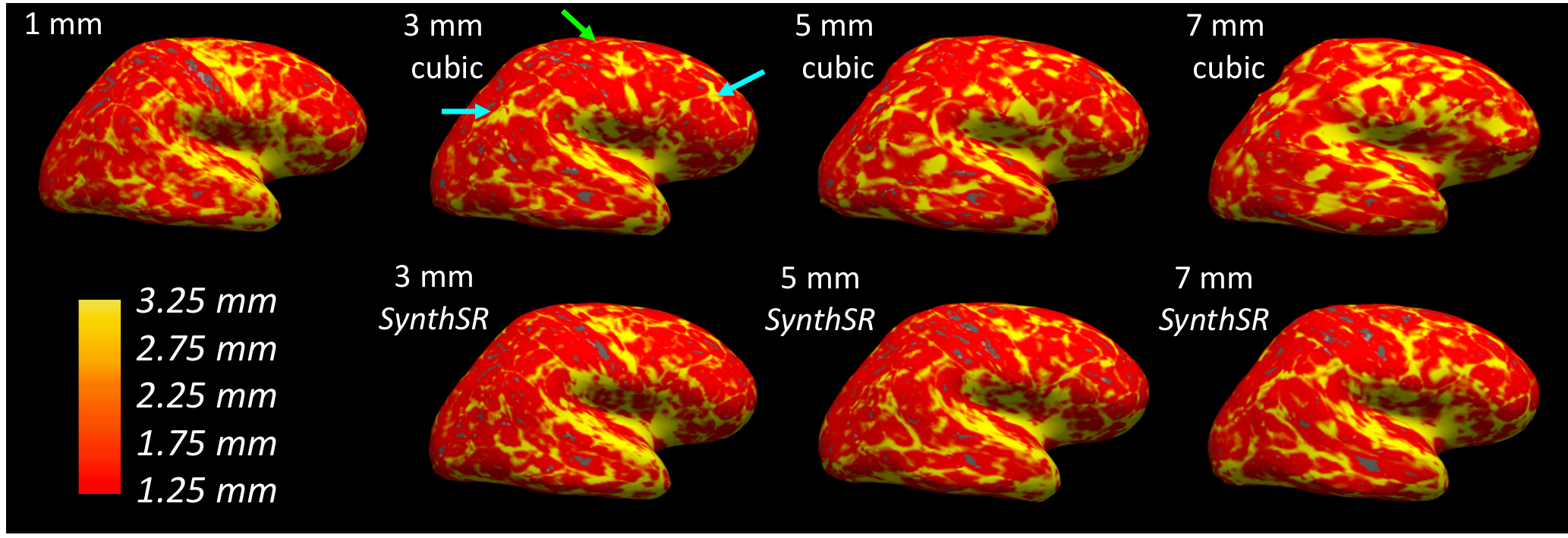}
\captionsetup{width=0.5\textwidth}
\caption{Thickness map for the right hemisphere of the subject in Figure~\ref{fig:upsampling}, derived from different slice thicknesses, with cubic interpolation and \emph{SynthSR}. The thickness maps are displayed on the inflated surface. The blue arrows point at regions of overestimated thickness (inferior parietal, rostral middle frontal), and the green arrow points at a region where the thickness in underestimated (precentral).}
\label{fig:thicknessCase}
\end{figure}

\begin{figure}[!t]
\centering
\includegraphics[width=0.5\textwidth, keepaspectratio]{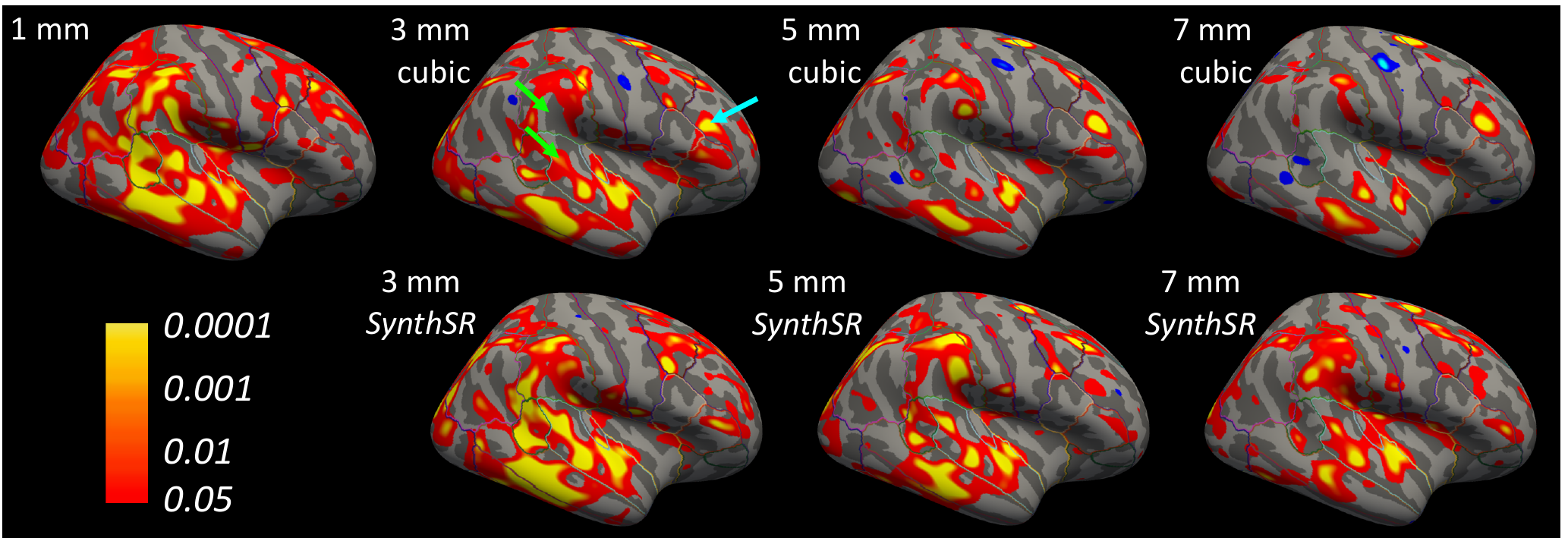}
\captionsetup{width=0.5\textwidth}
\caption{Significance maps (logarithmic scale) for AD vs controls in right hemisphere, corrected for age and sex, for different slice thicknesses. The results are displayed on the inflated surface of FreeSurfer’s template ``fsaverage''. The green arrows point at false negatives (supramarginal, superior temporal), and the blue arrow points at a false positive (rostral middle frontal).}
\label{fig:thicknessGroup}
\end{figure}

\paragraph{Cortical thickness}
We conducted a similar experiment with cortical thickness, where we compared the results when analyzing 3, 5 and 7 mm coronal scans with FreeSurfer 7, and the reference obtained by running FreeSurfer 7 on the original 1 mm scans. Figure~\ref{fig:thicknessCase}  shows the results for the right hemisphere of the subject in  Figure~\ref{fig:upsampling}, whereas Figure~\ref{fig:thicknessGroup} shows significance maps for the AD vs. controls test, correcting for age and sex. The isotropic 1 mm data show expected effects in the temporal and supramarginal regions \citep{lerch2005focal,querbes2009early,lehmann2011cortical,li2012discriminant}. 
 Cortical thickness is, as expected, more sensitive to insufficient resolution than subcortical volumetry. When cubic interpolation without SR is used, large errors render the data nearly useless already at 3 mm spacing, both for the individual (e.g., reduced thickness in precentral region, increased in inferior parietal and rostral middle frontal; see arrows in the figure) and the group study (e.g., false negatives in supramarginal and superior temporal; false positive in rostral middle frontal; see arrows).  SR with \emph{SynthSR}, on the other hand, yields maps that are very similar to the isotropic reference at 3 mm spacing. Many clusters persist even at 5 and 7 mm, albeit with reduced significance at the group level. A similar trend is observed when comparing the estimated area of the pial surface (Table~\ref{tab:surfaceArea}): without SR, many deeper sulci are missed, leading to greatly underestimated surface areas (7.7\% at 3 mm, 9.5\% at 5 mm, and 13.0\% at 7 mm). \emph{SynthSR} recovers large part of the lost surface area, especially at 3 mm and 5 mm resolution -- which is consistent with the other results in Figures~\ref{fig:thicknessCase} and~\ref{fig:thicknessGroup}.
  
\begin{figure}
\centering
\includegraphics[width=0.5\textwidth, keepaspectratio]{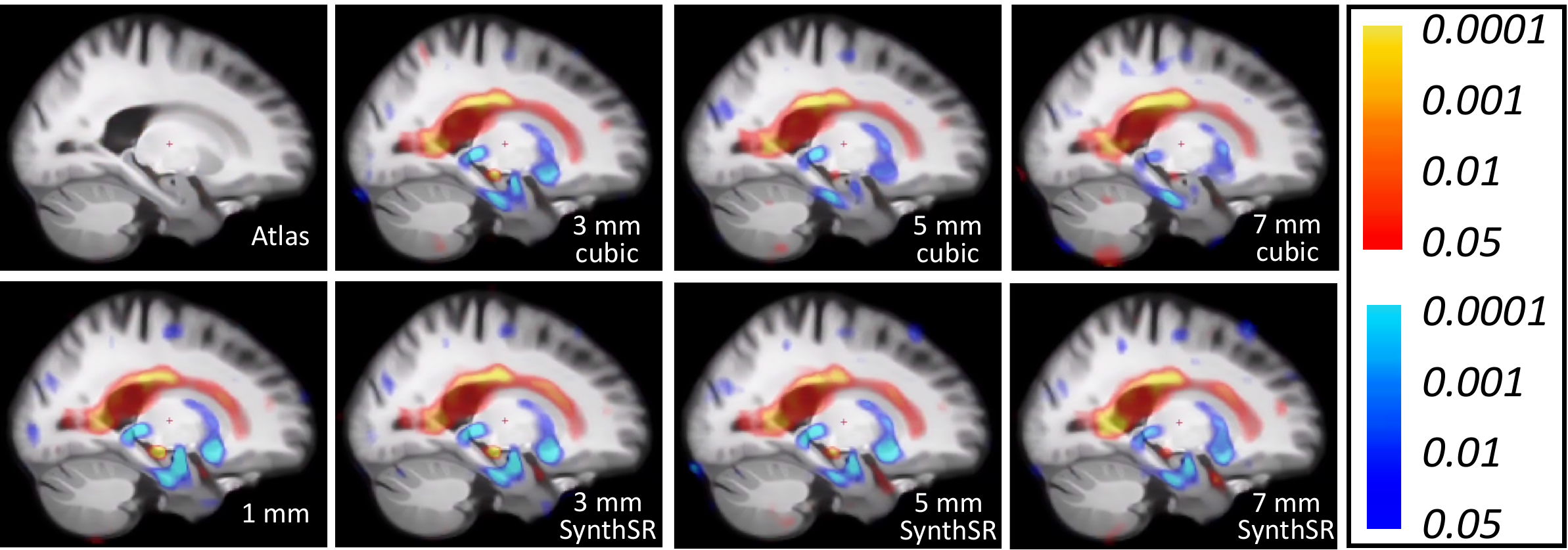}
\captionsetup{width=0.5\textwidth}
\caption{Significance maps of TBM of AD vs. controls at different resolutions, with and without SR (\emph{SynthSR}). Blue indicates more contraction in AD, and red indicates more expansion.}
\label{fig:TBM}
\end{figure}

\begin{table*}[!t]
\centering
 \begin{tabular}{|c|c|c|c|c|c|c|c|}
  \hline
 Resolution & 1 mm &  \makecell{ 3 mm \\ cubic } & \makecell{ 3 mm \\ \emph{SynthSR} } & \makecell{ 5 mm \\ cubic } & \makecell{ 5 mm \\ \emph{SynthSR} } &  \makecell{ 7 mm \\ cubic } & \makecell{ 7 mm \\ \emph{SynthSR} } \\
  \hline
 \makecell{ Average pial \\ surface area (mm$^2$)  } & \makecell{ 19,881 \\$\pm$ 1,876}&  \makecell{ 18,354 \\$\pm$  1,791}& \makecell{ 19,472 \\$\pm$ 1,850 }& \makecell{ 18,001 \\$\pm$  1,775}& \makecell{ 19,380 \\$\pm$ 1,943 }& \makecell{ 17,291 \\$\pm$ 1,708 }& \makecell{ 18,600 \\$\pm$ 1,809 }\\
  \hline
  \end{tabular}
\captionsetup{width=0.9\textwidth}
\caption{Average area of pial surface of subjects on the ADNI dataset, estimated with FreeSurfer 7 on scans of different coronal resolution, with and without super-resolution (i.e., cubic vs. \emph{SynthSR}).}
\label{tab:surfaceArea}
\end{table*}

\paragraph{Tensor-based morphometry}
In order to assess the usefulness of the \emph{SynthSR} volumes in registration, we investigated a TBM application \citep{freeborough1998modeling,chung2001unified,fox2001imaging,riddle2004characterizing} using a diffeomorphic registration algorithm with local normalized cross-correlation as similarity metric \citep{modat2010fast}. First, we computed a nonlinear atlas in an unbiased fashion (\citealt{joshi2004unbiased}, Figure~\ref{fig:TBM}, top left). Then, we compared the distribution of the Jacobian determinants between AD and controls, in atlas space, with a non-parametric Wilcoxon rank sum test. The results for the different resolutions are in the same figure.  The 1 mm isotropic volumes yield results that are consistent with the AD literature, e.g., contraction in the hippocampal head and tail as well as in the putamen, and expansion of ventricles \citep{hua2008tensor,de2008strongly,chupin2009fully}. Without SR (i.e., just cubic interpolation), significance already decreases noticeably at 3 mm spacing, and clusters disappear at 5 mm (e.g., hippocampal head, amygdala). Super-resolving with \emph{SynthSR}, all clusters still survive at 7 mm (with minimal loss of significance strength), indicating the power of the approach to accurately detect and quantify disease effects, even at large slice spacing.

\begin{figure}[!t]
\centering
\includegraphics[width=0.4\textwidth, keepaspectratio]{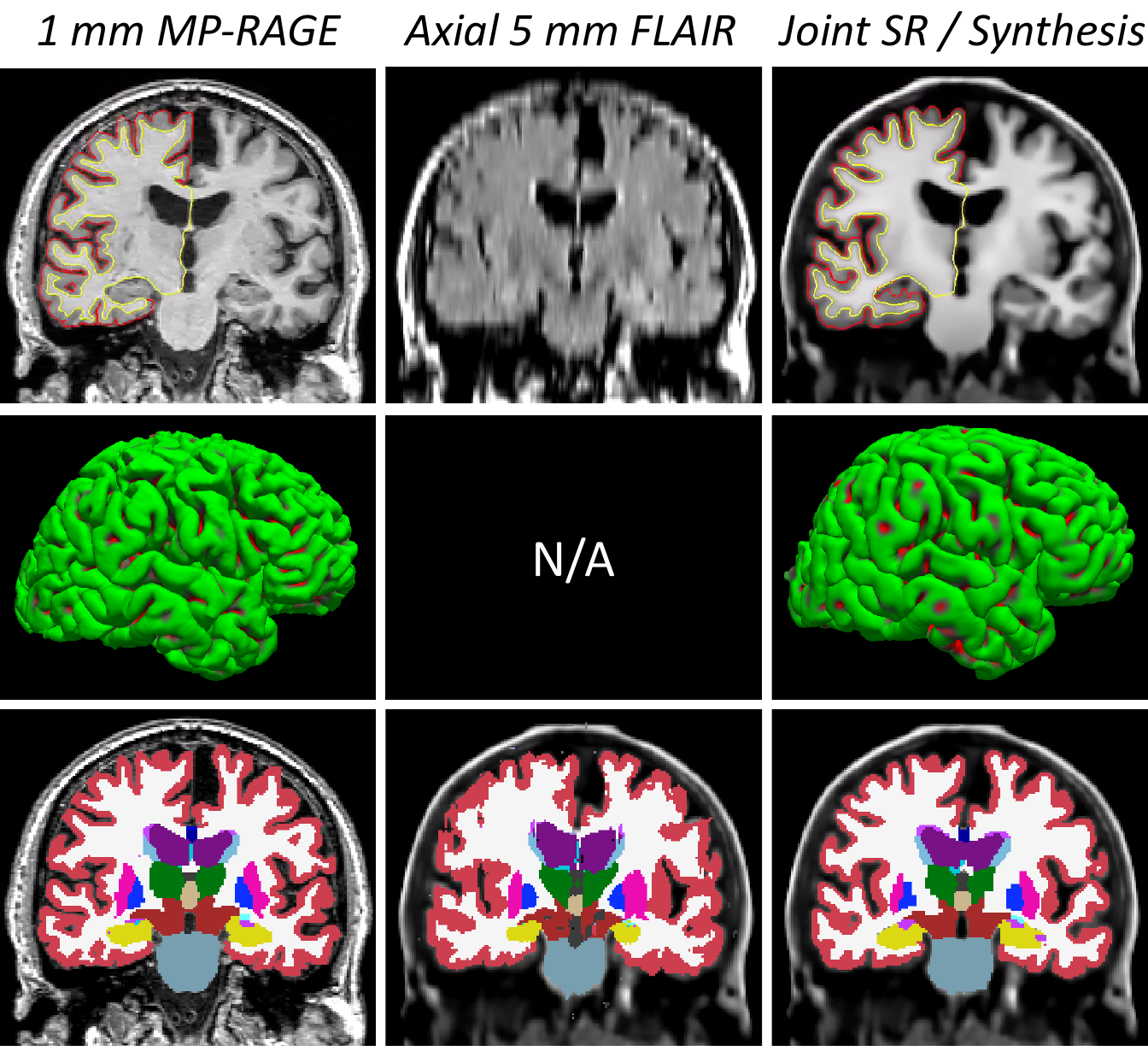}
\captionsetup{width=0.45\textwidth}
\caption{Coronal slice of a sample 1 mm T1 scan from the ADNI dataset (left); 5 mm axial FLAIR (with cubic interpolation; middle); and super-resolved, with \emph{SynthSR} (right). Top row: image intensities with pial and white matter surfaces of the right hemisphere computed with FreeSurfer 7 (not applicable to FLAIR scan). Middle row:  3D rendering of the pial surfaces. Bottom: volumetric segmentation obtained with FreeSurfer 7 (T1 and synthetic scans) and SAMSEG (FLAIR scan). Please note that the T1 and FLAIR scans are not perfectly aligned -- we did not register the T1 to avoid introducing interpolation artifacts that would propagate into the FreeSurfer processing.}
\label{fig:upsamplingFLAIR}
\end{figure}

\subsubsection{Joint super-resolution and synthesis of single, natively anisotropic  scans}

The second experiment assesses the performance the proposed method on a joint SR / synthesis problem using the FLAIR scans in ADNI, which were natively acquired at 5 mm spacing (and identical thickness) -- rather than artificially downsampled, as in the previous experiment. Working with ADNI scans has the advantage that we can use the measurements derived from the T1 scans as ground truth, as we did in the previous experiment. As opposed to the previous setup, we now use the real 1 mm scans as target, in order to produce synthetic scans of the reference T1 contrast, i.e., the MP-RAGE contrast of the training dataset.

\begin{figure*}[!t]
\centering
\includegraphics[width=0.9\textwidth, keepaspectratio]{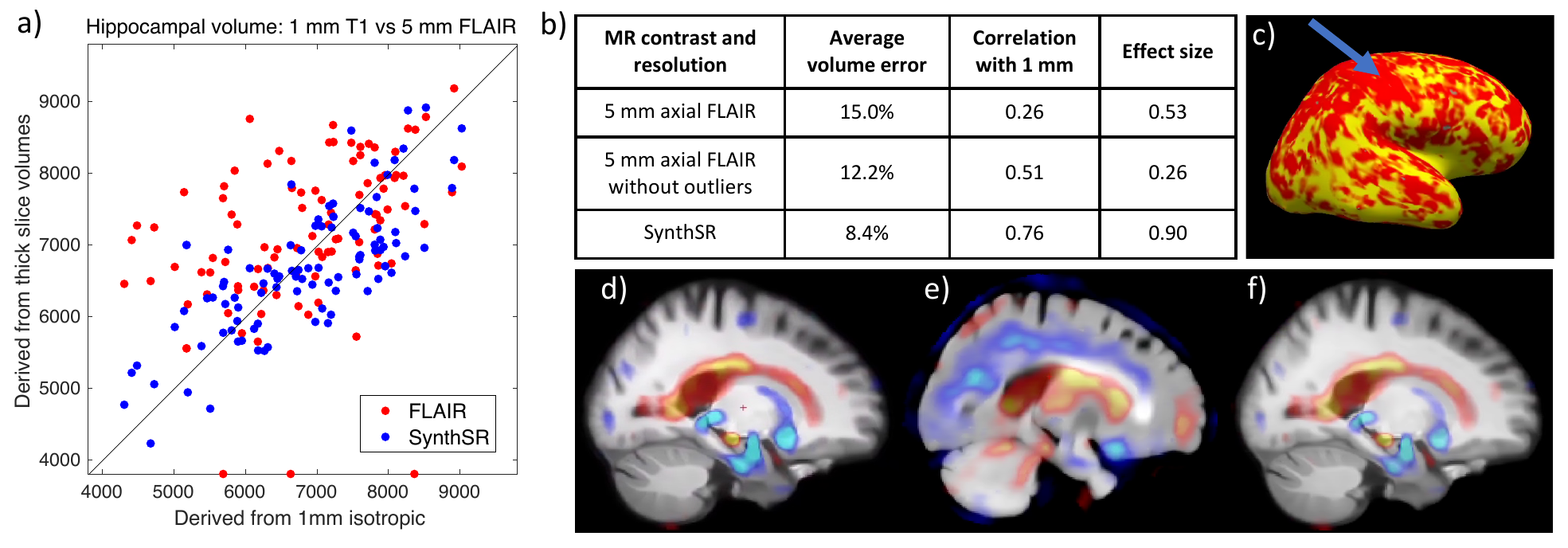}
\captionsetup{width=0.95\textwidth}
\caption{Summary of results for 5 mm axial FLAIR scans from ADNI; the ground truth is given by the measurements derived from the corresponding 1 mm MP-RAGE scans using FreeSurfer 7. (a)~Scatter plot for hippocampal volumes computed from  FLAIR scans (obtained with SAMSEG, using cubic interpolation) and from the 1 mm MP-RAGE scans produced by \emph{SynthSR} (obtained with FreeSurfer 7), against the ground truth volumes. The three dots on the $x$ axis represent outliers. (b)~Relative error in hippocampal volume, correlation with volumes from 1 mm T1 scans, and effect size of AD vs controls (corrected for sex, intracranial volume and age), for the 5 mm FLAIR scans (with and without considering the three outliers) and for the 1 mm MP-RAGE volumes produced by \emph{SynthSR}. (c)~Thickness map for the right hemisphere derived from the synthesized T1 scan of the same subjects as in Figure~\ref{fig:thicknessCase}. Compared with the ground truth in Figure~\ref{fig:thicknessCase} (top left), errors are rather noticeable (e.g., reduced thickness in the motor cortex, pointed by the arrow). (d-e)~TBM using the   ground truth T1 scans (d),   the 5 mm FLAIR scans (overlaid on its own FLAIR atlas, e), and the synthesized MP-RAGE volumes (f).}
\label{fig:FLAIRmisc}
\end{figure*}

Figure~\ref{fig:upsamplingFLAIR} shows an example of joint SR / synthesis for one of the FLAIR scans in the ADNI dataset. The limited gray / white matter contrast of the FLAIR input makes this task much more difficult than SR of MP-RAGE scans.  Nevertheless, \emph{SynthSR} is able to recover a very good approximation of the original volume, albeit smoother than in the previous experiment (e.g., Figure~\ref{fig:upsampling}). This smoothness leads to mistakes in the cortical segmentation, which, in spite of not appearing significant, have a large effect on cortical thickness estimation in relative terms (as shown by the results presented below), since the human cortex is only 2-3 mm thick on average. The subcortical structures, on the other hand, are a very good approximation to the ground truth obtained with the 1 mm MP-RAGE, and considerably better than the output produced by SAMSEG on the FLAIR scan upsampled with cubic interpolation, which has very visible problems -- including  poor cortical segmentation, largely oversegmented left putamen, or undersegmented hippocampi.

Figure~\ref{fig:FLAIRmisc} summarizes the results for the same hippocampal volumetry, cortical thickness and TBM analyses that we performed on the previous experiment. The hippocampal volumes (Figure~\ref{fig:FLAIRmisc}a) are more spread than when doing SR alone, but are still strongly correlated with the ground truth values, particularly considering two factors: the axial acquisition (much less suitable for imaging the hippocampus than the coronal plane) and the limited contrast that the hippocampus in FLAIR. These two aspects clearly deteriorate the performance of SAMSEG, which makes much larger errors (including three outliers where the hippocampus was largely undersegmented), particularly for subjects with more severe atrophy. This is reflected in the quantitative results in Figure~\ref{fig:FLAIRmisc}(b): even when the outliers are disregarded, the average volume error is over 12\%,  the correlation is only $\rho=0.51$, and effect size is barely 0.26. These values greatly improve to 8.4\% (volume error), $\rho=0.76$ (correlation) and 0.90 (effect size) respectively, when using the  1 mm T1 scans produced by \emph{SynthSR}. 

The cortical thickness maps are unfortunately not usable for this combination of contrast and resolution. Figure~\ref{fig:FLAIRmisc}c shows the thickness map of the subject from Figure~\ref{fig:upsamplingFLAIR}, which has obvious problems, e.g., it misses the expected, highly characteristic patterns in the precentral and postcentral cortices (pointed by the arrow; please compare with the 1 mm case in Figure~\ref{fig:thicknessCase}). Registration is, on the other hand, highly successful with \emph{SynthSR}: the TBM results (Figure~\ref{fig:FLAIRmisc}d) are nearly identical to those obtained with the real 1 mm T1 scans (Figure~\ref{fig:FLAIRmisc}e), whereas using the FLAIR scans directly (with a recomputed FLAIR atlas) leads to a large number of false negatives and positives.

\begin{figure}[!t]
\centering
\includegraphics[width=0.4\textwidth, keepaspectratio]{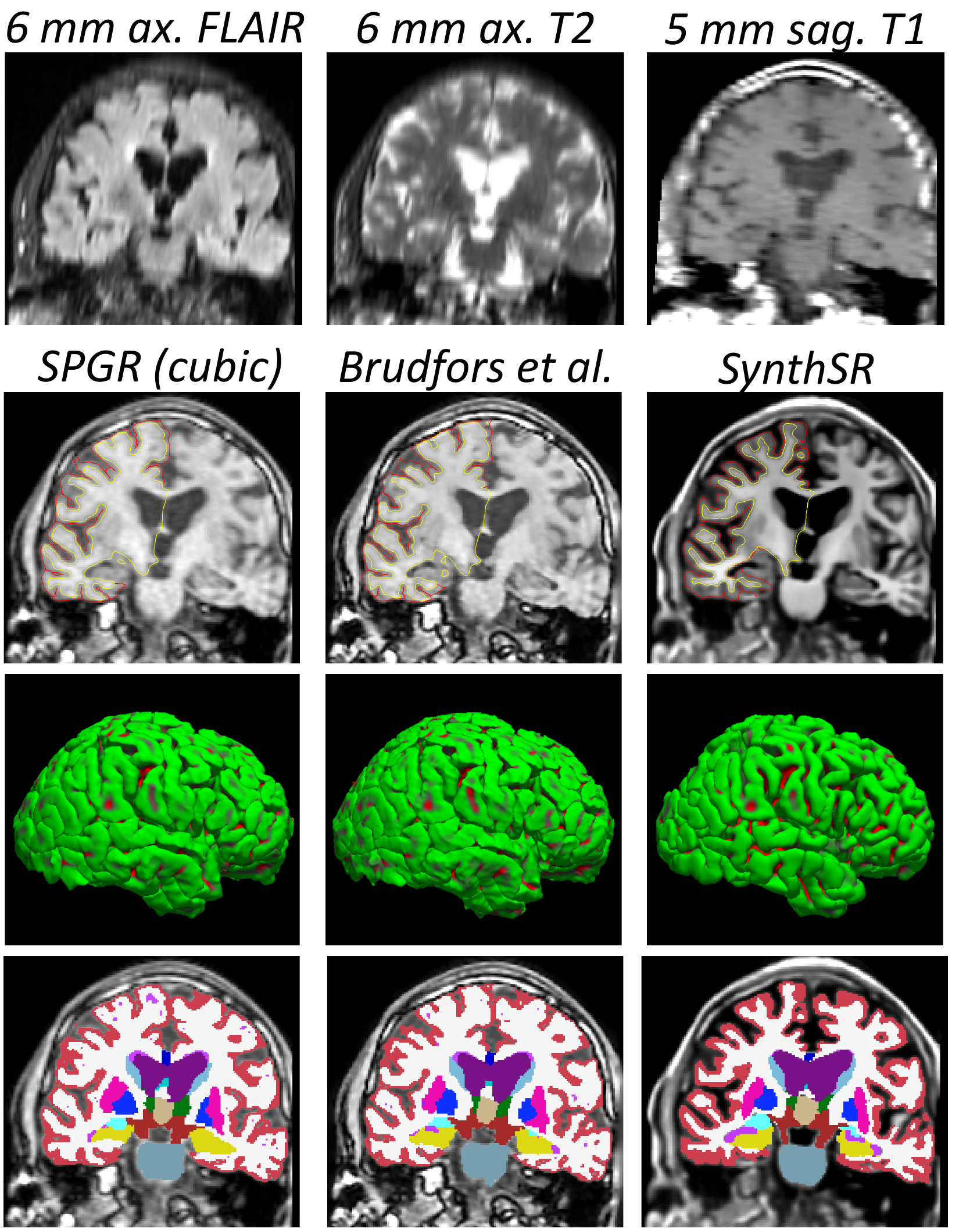}
\captionsetup{width=0.45\textwidth}
\caption{Joint SR / synthesis of an exam from the MGH dataset. The top row shows a coronal slice for the FLAIR, T2 and T1-TSE sequences, with cubic interpolation. The second row shows the corresponding T1-SPGR slice, along with the SR volume produced by \cite{brudfors2018mri}
 and the output from our method, with the pial and white matter surfaces of the right hemisphere computed with FreeSurfer 7. The third row shows the 3D rendering of the pial surfaces. The bottom row shows the  volumetric segmentation obtained with FreeSurfer 7.}
\label{fig:upsamplingConklin}
\end{figure}

\subsubsection{Super-resolution of clinical exams with multiple scans}

In this final experiment, we use the MGH dataset to evaluate \emph{SynthSR} in the scenario it was ultimately conceived for: joint SR and synthesis on multi-modal scans with channels of different resolution and MR contrast. We use the SPGR scan as reference (i.e., register the other scans to it), and then use \emph{SynthSR} to predict the residual between the upscaled SPGR and the desired MP-RAGE output, whose contrast is defined by the training dataset. Since there is no ground truth available for this dataset, we use qualitative evaluation, as well as indirect quantitative evaluation via an aging experiment; we note that we discarded three of the 40 cases, for which FreeSurfer completely failed to segment the SPGR scan.

\begin{figure*}[!t]
\centering
\includegraphics[width=0.90\textwidth, keepaspectratio]{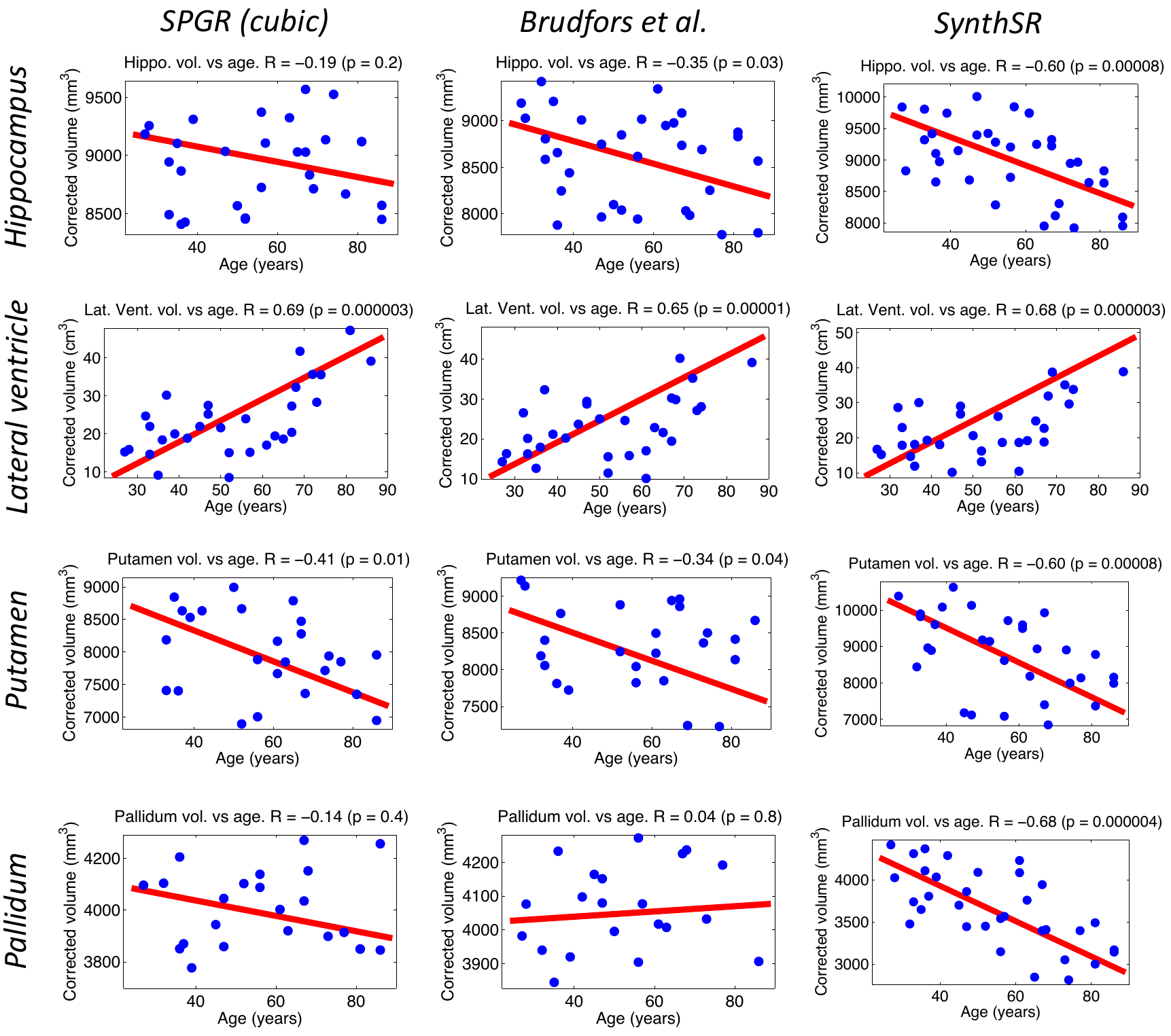}
\captionsetup{width=0.90\textwidth}
\caption{Scatter plots and linear regression of the bilateral volumes of the hippocampus, lateral ventricle and basal ganglia structures (putamen, pallidum) against age in the MGH dataset. The volumes were computed with FreeSurfer 7 from the SPGR scans directly (with cubic interpolation, left), their SR version produced by \citealt{brudfors2018mri} (middle), and the scans obtained with the joint SR / synthesis version of \emph{SynthSR} (right). The volumes are corrected by sex and intracranial volume. The correlation coefficients and the p value for their significance are shown in the title of each plot.}
\label{fig:ConklinVolumes}
\end{figure*}

Figure~\ref{fig:upsamplingConklin} shows an example from the MGH dataset. Directly using the low-quality SPGR with cubic interpolation has numerous problems.  Cortically, the lack of image contrast leads to poorly fitted surfaces that frequently leak into the dura matter, leading to unnaturally flat pial surfaces. Subcortically, PV and the overall lack of contrast force the FreeSurfer segmentation algorithm to heavily trust the prior; the example in the figure illustrates this problem well in the hippocampus (yellow) and the basal ganglia (putamen and pallidum, in pink and dark blue, respectively). The ability of the SPRG scans to capture well-known age effects \citep{potvin2016normative} is considerably hampered by these segmentation mistakes (Figure~\ref{fig:ConklinVolumes}): while very obvious large-scale features like ventricular expansion are accurately detected (even with its characteristic quadratic shape), the atrophy of the hippocampus and basal ganglia (correcting for sex and intracranial volume) are mostly missed (only the putamen is borderline significant). \citeauthor{brudfors2018mri}'s method exploit the information on the other scans to achieve some sharpening that moderately improves the subcortical segmentation (e.g., improves the correlation of hippocampal volume and age), while having very little effect on the placement of cortical surfaces. 

Conversely,  \emph{SynthSR} yields much better contrast between gray and white matter, as well as crisper boundaries. This enhanced image quality enables FreeSurfer to generate more plausible cortical surfaces, as well as a much more precise segmentation of subcortical structures (e.g., the basal ganglia or the hippocampi in Figure~\ref{fig:upsamplingConklin}). This superior contrast is also reflected in the aging analysis: the volumes computed with FreeSurfer on the scans obtained with \emph{SynthSR} successfully detect all the expected effects, i.e., atrophy of the hippocampus and basal ganglia and expansion of the lateral ventricles. The improvement with respect to  \citeauthor{brudfors2018mri}'s method is  very clear: \emph{SynthSR} detects the negative slope with p$<$10$^{-4}$ for all structures, whereas  their approach yields p$>$10$^{-2}$ in all cases (lateral ventricle aside), and is completely unable to detect the slope effect in the pallidum, despite the fair sample size (37 subjects).

\section{Discussion and conclusion}
\label{sec:discussion}

In this article, we have presented \emph{SynthSR}, the first learning method that produces an isotropic volume of reference MR contrast using a set of scans from a routine clinical MRI exam consisting of anisotropic 2D acquisitions, without access to high-resolution training data for the input modalities.  \emph{SynthSR} uses  random synthetic data mimicking the resolution and contrast of the scans one aims to super-resolve, to train a regression CNN that produces the desired HR intensities with the target contrast. The synthetic data are generated on the GPU on the fly with a mechanism inspired by the generative model of Bayesian segmentation, which enables simulation not only of contrast and resolution, but also changes in orientation, subject motion between scans, as bias field and registration errors. Because such artifacts and extracerebral tissue are included in the simulations, our method does not require any preprocessing (e.g., skull stripping, denoising, or bias field correction).

The first set of experiments on SR alone reveals that \emph{SynthSR} can super-resolve MRI scans very accurately, despite the domain gap between real and synthetic data. Using artificially downsampled MP-RAGE scans from ADNI shows that one can replace 1 mm isotropic scans by super-resolved acquisitions of much lower native resolution and still detect the expected effects of disease. Our results show that, in the context of registration and subcortical segmentation, one can go down to 5 or even 7 mm slice spacing without almost any noticeable impact on common downstream analyses. Cortical thickness is, as expected, much more sensitive to larger spacing, but the proposed technique enables reliable thickness analysis at 3 mm spacing -- which is remarkable, given the convoluted shape of the cortex and the small size of the thinning effect one seeks to detect. 

When SR and synthesis are combined, the problem becomes much harder. Our experiments with 5 mm FLAIR scans show that cortical thickness analysis on the synthesized 1 mm MP-RAGE volumes is not reliable. Moreover, the subcortical segmentations produce volumes that yield lower effect sizes and correlations with the ground truth than when performing SR of T1 scans.  However, the hippocampal volumes obtained with \emph{SynthSR} are still usable, in absolute terms (their correlation with the ground truth volumes is over 0.75). This result is noteworthy, particularly given the axial orientation of the FLAIR scans, which is approximately parallel to the major axis of the hippocampus -- causing a very robust tool like SAMSEG to visibly falter.

The results on the MGH dataset show that \emph{SynthSR} can effectively exploit images with different contrast and orientation. Compared with the outputs from the second experiment, the synthetic 1 mm MP-RAGEs have much better contrast in regions where it is difficult to define boundaries from a FLAIR scan alone -- compare, for instance, the contrast of the putamen in Figures~\ref{fig:upsamplingFLAIR} and \ref{fig:upsamplingConklin}. Even though obvious effects like ventricular expansion can be measured even with lower-resolution scans, the superior image quality produced by our approach enables FreeSurfer to reproduce subtler signatures of aging that are missed by the competing approach (e.g., pallidum). Unfortunately, as with the FLAIR scans from ADNI, the image quality of this dataset was insufficient for our method to accurately detect expected patterns of aging in cortical thickness.

We emphasize that it is not the goal of this work to replace image acquisition for a single specific subject. Rather, our goal is to enable analyses with existing neuroimaging tools that are not otherwise possible with the thick-slice scans that are used in a majority of routine clinical brain MRI protocols. Our results show that  isotropic scans synthesized with \emph{SynthSR} can be used to compute good registrations and segmentations in many cases, almost as good as the real 1 mm scans in many analyses at the group level. Even though analysis like atrophy estimation via longitudinal segmentation or registration using the synthetic scans may be informative to evaluate a patient in clinical practice, we do not envision our method replacing specific MRI acquisitions (e.g., with contrast agents) for evaluation of abnormalities like tumors.

While it is not the goal to produce harmonized data for multi-center studies,  \emph{SynthSR}  generates synthetic scans of a specific predefined MR contrast. Although this indirectly achieves a level of harmonization, it does not homogenize the data as well as explicit harmonization techniques: with \emph{SynthSR}, the ability to generate contrast in the output depends on the quality and contrast of the input scans (e.g., as in the aforementioned example of the putamen in Figures~\ref{fig:upsamplingFLAIR} and \ref{fig:upsamplingConklin}). It may thus be interesting to build a pipeline with our method and existing harmonization methods (e.g.,  \citealt{pomponio2020harmonization}), possibly within a single architecture trained end to end. 

Further work will be directed towards improving the robustness and accuracy of the approach presented in this article, ideally to the point that cortical thickness analyses are possible. Improving our method is possible in many aspects. In terms of loss, one could replace L1 by adversarial networks that seek to make the generated volumes indistinguishable from the training scans. While this approach generates very realistic images, it is also more prone to hallucinating image features \citep{cohen2018distribution}. Therefore, it will be important to compare the performance in downstream analyses. A simpler alternative may be to produce more realistic synthetic images in training by using finer labels. Crucially, labels do not need to be manual or correspond one-to-one with structures: since they are not used in learning (as opposed to, e.g., a segmentation problem), they can be obtained in an automated fashion, e.g., with unsupervised clustering techniques like \cite{blaiotta2018generative}. 

Further improvements to \emph{SynthSR} are also possible in terms of architecture. While the U-net in this paper has been successfully applied to a number of related problems, deep machine learning advances at great speed, so it is almost certain that improved results will be obtained with more modern architectures in the future. We will also attempt to improve the image augmentation model. When deploying our method on clinical data at larger scale, the CNN will encounter images with higher degrees of noise and motion than the relatively small MGH dataset used in this study. Incorporating these artifacts into our augmentation model may improve the results. When testing at scale, we expect that some MR modalities from our minimal subset (FLAIR, T1-TSE, T2, SPGR) will be missing or unusable. While this could be addressed by training a CNN for every possible subset, we will also try training a single CNN with modality dropout. Such a CNN could potentially be applied to any MRI exam, irrespective of what modalities are available. This approach would also require the ability to automatically determine what scans within an exam are usable, which is a challenge of its own.  

Finally, a crucial development that is required to run \emph{SynthSR} at scale in the clinic is the ability to model pathology. While effects like atrophy can largely be captured with spatial augmentation, coping with more structurally disrupting abnormalities (e.g., tumors) will require simulating them in training. Given that \emph{SynthSR} seems to be able to cope with a fair amount of domain gap between synthetic and real intensities, it is unclear how accurate these simulations will have to be.

\emph{SynthSR} is publicly available (at \url{https://github.com/BBillot/SynthSR}) and will enable researchers around the globe to generate synthetic 1 mm scans from vast amounts of brain MRI data that already exist and are continuously being acquired. These synthetic scans will enable the application of many existing neuroimaging tools designed for research-grade MRI (including but not limited to the ones in this paper) to huge sample sizes, and thus hold promise to improve our understanding of the human brain by providing levels of statistical power that are currently not attainable with \emph{in vivo} studies.

\section*{Acknowledgement}

This project has been primarily funded by the European Research Council (Starting Grant 677697, project ``BUNGEE-TOOLS''), the NIH (1RF1-MH-123195-01), and Alzheimers Research UK (Interdisciplinary Grant ARUK-IRG2019A-003). Further support has been provided by the EPSRC (EP-L016478-1,  EP-M020533-1, EP-R014019-1), the NIHR UCLH Biomedical Research Centre, the BRAIN Initiative Cell Census Network (U01-MH117023), the National Institute for Biomedical Imaging and Bioengineering (P41-EB-015896, 1R01-EB-023281,
R01-EB-006758, R21-EB-018907, R01-EB-019956, P41-EB-015902), the National Institute on
Aging (1R56-AG064027, 1R01-AG064027, 5R01-AG008122, R01-AG016495), the National
Institute of Mental Health, the National Institute of Child Health and Human Development (R01-HD100009), the National Institute of Diabetes and
Digestive and Kidney Diseases (R21-DK108277-01), the National
Institute for Neurological Disorders and Stroke (R01-NS-0525851, R21-NS-072652,
R01-NS-070963, R01-NS-083534, 5U01-NS-086625, 5U24-NS-10059103, R01-NS-105820, R21-NS-109627, RF1-NS-115268, U19-NS-115388), NIH Director's Office (DP2-HD-101400), James S. McDonnell Foundation, the Tiny Blue Dot Foundation,
and was made possible by the resources provided by Shared
Instrumentation Grants 1S10-RR023401, 1S10-RR019307, and
1S10-RR023043. Additional support was provided by the NIH Blueprint for
Neuroscience Research (5U01-MH093765), part of the multi-institutional
Human Connectome Project. In addition, BF has a financial interest in
CorticoMetrics, a company whose medical pursuits focus on brain
imaging and measurement technologies. BF's interests were reviewed and
are managed by Massachusetts General Hospital and Partners 
HealthCare in accordance with their conflict of interest policies.

The collection and sharing of the MRI data used in the group study based on ADNI was funded by the Alzheimer's Disease Neuroimaging Initiative (NIH grant U01-AG024904) and DOD ADNI (Department of Defence award number W81XWH-12-2-0012). ADNI is funded by the National Institute on Aging, the National Institute of Biomedical Imaging and Bioengineering, and through generous contributions from the following: Alzheimer's Association; Alzheimer's Drug Discovery Foundation; BioClinica, Inc.; Biogen Idec Inc.; Bristol-Myers Squibb Company; Eisai Inc.; Elan Pharmaceuticals, Inc.; Eli Lilly and Company; F. Hoffmann-La Roche Ltd and its affiliated company Genentech, Inc.; GE Healthcare; Innogenetics, N.V.; IXICO Ltd.; Janssen Alzheimer Immunotherapy Research \& Development, LLC.; Johnson \& Johnson Pharmaceutical Research \& Development LLC.; Medpace, Inc.; Merck \& Co., Inc.; Meso Scale Diagnostics, LLC.; NeuroRx Research; Novartis Pharmaceuticals Corporation; Pfizer Inc.; Piramal Imaging; Servier; Synarc Inc.; and Takeda Pharmaceutical Company. The Canadian Institutes of Health Research is providing funds to support ADNI clinical sites in Canada. Private sector contributions are facilitated by the Foundation for the National Institutes of Health (www.fnih.org). The grantee organisation is the Northern California Institute for Research and Education, and the study is coordinated by the Alzheimer's Disease Cooperative Study at the University of California, San Diego. ADNI data are disseminated by the Laboratory for Neuro Imaging at the University of Southern California.

\bibliographystyle{model5-names}\biboptions{authoryear}

\bibliography{biblio}

\end{document}